\documentclass[10pt, twocolumn]{extarticle}
\usepackage{graphicx}
\usepackage{hyperref}
\usepackage{amssymb}
\usepackage{amsmath}
\usepackage{amsfonts}
\usepackage{parskip}
\usepackage{titling}
\usepackage[table]{xcolor}
\usepackage[square,sort&compress,comma,numbers]{natbib}
\usepackage{multirow}
\usepackage{empheq}
\usepackage{caption}
\usepackage{multicol}

\usepackage{titlesec}
\usepackage{lipsum}
\titleformat{\section}{\large\bf}{\thesection}{1em}{}
\titleformat{\subsection}{\bf}{\thesubsection}{1em}{}
\titleformat{\subsubsection}{\it}{\thesubsubsection}{1em}{}
\pretitle{\Large\bf\vskip 0.3em} \posttitle{\par\vskip 0.6em} \preauthor{\large} \postauthor{\par\vskip -0.0em}
\predate{\large} \postdate{\par\vskip 0.1em \rule{\linewidth}{0.5pt}} 

\newcommand{\beq}{\begin{equation}}
\newcommand{\eeq}{\end{equation}}
\newcommand{\bea}{\begin{eqnarray}}
\newcommand{\eea}{\end{eqnarray}}

\newcommand{\comment}[1]{}
\renewcommand{\d}{{\rm d}}

\newcommand{\SuppI}{\ref{sec:si-est}}
\newcommand{\SuppII}{\ref{sec:si-imp}}
\newcommand{\SuppIII}{\ref{sec:si-ecm}}

\begin{document}

\title{Asymptotically Fault-Tolerant Programmable Photonics}
\author{Ryan Hamerly$^{1,2,*}$, Saumil Bandyopadhyay$^1$, Dirk Englund$^1$}
\date{November 29, 2022}


\maketitle



\begin{flushleft}
\small
$^{1}$      \textit{Research Laboratory of Electronics, MIT, 50 Vassar Street, Cambridge, MA 02139, USA} \\
$^{2}$      \textit{NTT Research Inc., Physics and Informatics Laboratories, 940 Stewart Drive, Sunnyvale, CA 94085, USA} \\
$^{*}$      rhamerly@mit.edu
\end{flushleft}

{\bf\small Abstract---Component errors limit the scaling of programmable coherent photonic circuits.  These errors arise because the standard tunable photonic coupler---the Mach-Zehnder interferometer (MZI)---cannot be perfectly programmed to the cross state.  Here, we introduce two modified circuit architectures that overcome this limitation: (1) a 3-splitter MZI mesh for generic errors, and (2) a broadband MZI+Crossing design for correlated errors.  Because these designs allow for perfect realization of the cross state, the matrix fidelity no longer decreases with mesh size, allowing scaling to arbitrarily large meshes.  The proposed architectures support progressive self-configuration, are more compact than previous MZI-doubling schemes, and do not require additional phase shifters.  This eliminates a major obstacle to the development of very-large-scale linear photonic circuits.}

\rule{\linewidth}{0.5pt}

\begin{figure}[b!]
\begin{center}
\includegraphics[width=1.00\columnwidth]{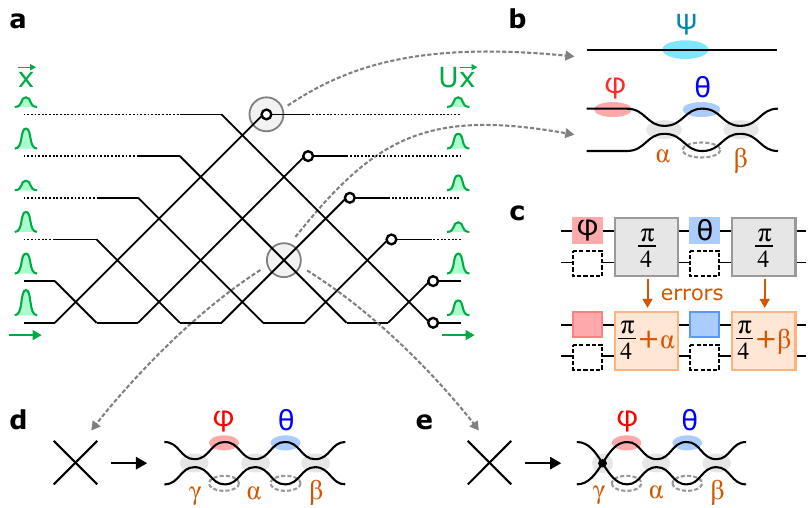}
\caption{Multiport interferometers with imperfect components.  (a) $6\times 6$ triangular mesh, composed of (b) a phase screen $\psi$ and tunable MZI couplers $\theta, \phi$.  (c) Fabrication imperfections lead to splitting-ratio errors $\alpha, \beta$.  (d-e) Alternative error-resilient coupler designs proposed in this paper: (d) 3-splitter MZI and (e) MZI+Crossing.}
\label{fig:f1}
\end{center}
\end{figure}

Large-scale programmable photonic circuits are opening up radical new possibilities for optics.  Of central importance in many devices is the universal multiport interferometer, which functions as an $N\times N$ programmable linear circuit (Fig.~\ref{fig:f1}(a-b)).  This device, usually constructed from a dense mesh of Mach-Zehnder interferometers (MZIs) \cite{Reck1994, Clements2016}, is widely employed in applications ranging from spatially multiplexed optical communications to machine learning and quantum computing \cite{Carolan2015, Zhong2020, Shen2017, Marpaung2013, Zhuang2015}.  Sadly, component errors (Fig.~\ref{fig:f1}(c)) are a critical factor limiting the size of such circuits.  Since the circuit depth of MZI meshes scales as $O(N)$, the effect of errors grows with mesh size, meaning that, in practice, even modestly sized circuits cannot be programmed to high accuracy.  Motivated by this challenge, a large body of recent work has focused on ``correcting'' hardware errors by global optimization \cite{Burgwal2017, Mower2015, Pai2019}, self-configuration \cite{Pai2020, Hughes2018, Miller2013a, Miller2013b, Miller2017, RyanPaper1, RyanPaper2, Annoni2017}, or local correction \cite{Bandyopadhyay2021, Kumar2021}.  For conventional MZI meshes, correction reduces errors by a quadratic factor \cite{Bandyopadhyay2021, RyanPaper1}; however, the effect of errors still grows with mesh size and poses a fundamental limit to the scaling of these circuits.

To overcome this limit, various alternative mesh architectures have been proposed.  Non-compact structures such as binary trees avoid the extreme splitting-ratio requirements \cite{Lopez2019, JasvithPaper}, but suffer from large chip area and the need for many crossings.  A complementary approach is to stick to conventional geometries \cite{Reck1994, Clements2016}, but insert redundant MZIs to realize the full range of splitting ratios even in imperfect hardware \cite{Miller2015, Suzuki2015, Wilkes2016}.  This solves the scaling problem, but at the cost of a 1.5--2$\times$ increase in the number of splitters and phase shifters.  The resulting effects on chip area (particularly on emerging high-speed platforms where phase shifters have a large footprint \cite{Wu2019, Dong2022}), waveguide length (which affects insertion loss and latency \cite{Bandyopadhyay2022}), and electronic complexity (number of pads, traces, DACs / drivers, etc.) make this option unappealing.

In this paper, we propose two mesh architectures that achieve the same perfect scaling without significant added complexity: a 3-splitter MZI that corrects all hardware errors (Fig.~\ref{fig:f1}(d)) and an MZI+Crossing design that only corrects correlated errors, but has the added advantage of broader bandwidth (Fig.~\ref{fig:f1}(e)).  These designs take up significantly less chip area than the ``perfect'' redundant MZIs \cite{Miller2015, Suzuki2015}, and do not require additional phase shifters.  Moreover, the proposed architectures support progressive self-configuration \cite{RyanPaper1, RyanPaper2}, allowing for error correction even when the hardware errors are unknown.  This work will enable the development of freely scalable, broadband, and compact linear photonic circuits.

This paper is structured as follows: first we introduce the formalism of error correction in MZI meshes, focusing on the self-configuration approach.  Splitting ratios are visualized as points on the Riemann sphere, where hardware imperfections lead to forbidden regions around the poles (bar- and cross-state), where the probability density is at a maximum.  To avoid this unfortunate coincidence, our architectures ``rotate'' the Riemann sphere to move the forbidden regions away from this peak, so that a larger fraction of MZIs are perfectly realized.  Based on this concept, we introduce the 3-splitter MZI, which can correct arbitrary errors by rotating the forbidden regions to the equator.  Using a benchmark optical neural network, we show that this modified MZI mesh is $> 3\times$ more robust to hardware errors, enabling accurate inference in a regime where standard interferometric circuits struggle.  Finally, we introduce the MZI+Crossing, which flips the poles of the Riemann sphere.  While this design is only robust against correlated errors, it has the added advantage of broader intrinsic bandwidth.  For both architectures, we compare the matrix fidelity to the standard MZI to demonstrate the scaling advantage of both schemes.


\section*{Results}
\subsection*{Error Correction Formalism}
%

\begin{figure}[tb]
\begin{center}
\includegraphics[width=1.00\columnwidth]{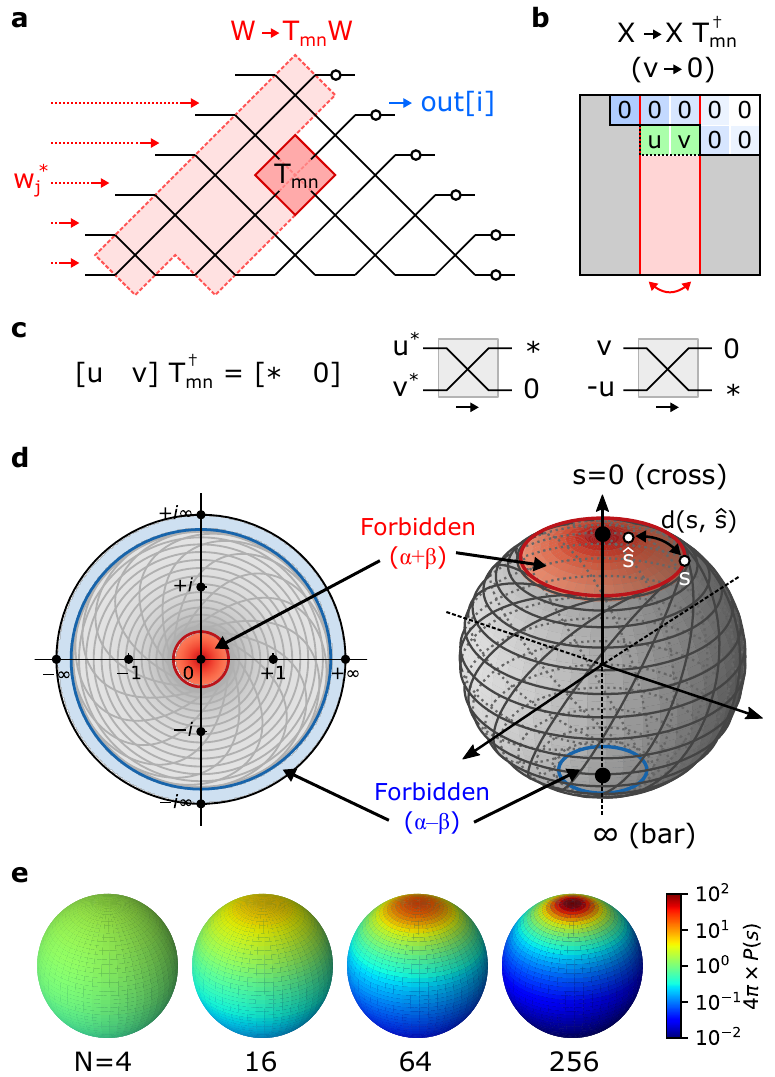}
\caption{Nulling method of self configuration.  (a) Configuring MZI $T_{mn}$ updates matrix $W$.  (b) Corresponding nulling update to $X = U W^\dagger$, which is (c) equivalent to zeroing an output of $T_{mn}$ given a fixed input.  (d) Allowed range of $s = T_{11}/T_{12} \in \mathbb{C}$; regions near $s = 0$ and $s = \infty$ are forbidden due to imperfections.  Contours are lines of constant $(\theta, \phi)$, with $\alpha = 0.23, \beta = 0.07$.  (e) Probability density $P(s)$ as a function of mesh size.}
\label{fig:f2}
\end{center}
\end{figure}

To correctly configure an MZI mesh in the presence of errors, one uses a {\it nulling} method based on physical measurements \cite{RyanPaper1, RyanPaper2}.  Fig.~\ref{fig:f2}(a) illustrates the case of the triangular mesh \cite{Reck1994}, where the procedure is more straightforward.  The transfer matrix for this system is a product of a phase screen $D$ and a sequence of $2\times 2$ unitaries $W$:
\beq
	U = D \underbrace{\prod_{mn} T_{mn}}_{W}
\eeq
where $T_{mn}$ is the $n^{\rm th}$ MZI of the $m^{\rm th}$ rising diagonal.
We configure the mesh by building up matrix $W$ in a sequence of steps designed to diagonalize a target matrix $X = U W^\dagger$.  In each step, we add one crossing to $W$, performing the update $W \rightarrow T_{mn} W$ 
, which right-multiplies the target matrix $X \rightarrow X T_{mn}^\dagger$ 
(Fig.~\ref{fig:f2}(b)).  The phase shifts $(\theta, \phi)$ are chosen to zero a particular matrix element $v \rightarrow 0$ (green in figure), satisfying the equation (indices $m, n$ suppressed for notational simplicity):
\beq
	[u\ \ v] T^\dagger = [*\ \ 0] \ \ \ \Leftrightarrow\ \ \ {T_{11}}/{T_{12}} = {u}/{v} \label{eq:ii}
\eeq
This is illustrated in Fig.~\ref{fig:f2}(c).  Nulling physically corresponds to injecting $w_j^*$ (the $j^{\rm th}$ column of $W^\dagger$) and zeroing the power at the $i^{\rm th}$ output \cite{RyanPaper2}.  If all nulling steps are performed exactly, the mesh will perfectly realize the target matrix $U$ (see Methods and Supp.~Sec.~\SuppIII~for details).

Mathematically, nulling corresponds to matching the complex splitting ratio $s \equiv T_{11}/T_{12} = -(T_{22}/T_{21})^*$ to a target value $\hat{s} \equiv u/v$.  This is not always possible, as the range of splitting ratios $\tan |\alpha + \beta| \leq |s| \leq \cot |\alpha - \beta|$ is constrained by hardware imperfections, namely the splitting-angle errors $\alpha, \beta$ for the 50:50 couplers in a real MZI (Fig.~\ref{fig:f1}(c)).  These imperfections lead to {\it forbidden regions} (Fig.~\ref{fig:f2}(d)) for small and large $s$, where nulling cannot be achieved perfectly.  It is also instructive to view this chart on the Riemann sphere, which shows that these forbidden regions are centered around the poles (Fig.~\ref{fig:f2}(b)), highlighting the well-known fact that imperfect MZIs generally have finite extinction ratio and cannot realize a perfect cross ($s = 0$) or bar ($s = \infty$) state.
If in a given nulling step $\hat{s}$ falls within the forbidden region, nulling is imperfect, and an off-diagonal residual prevents perfect diagonalization of the matrix, leading to an ``uncorrectable'' error.  This residual is proportional to $d(s, \hat{s})$, the Euclidean distance on the Riemann sphere between the target ratio and the closest realizable $s$.  The overall error 
is the quadrature sum of all such residuals.

\begin{figure*}[t!]
\begin{center}
\includegraphics[width=1.00\textwidth]{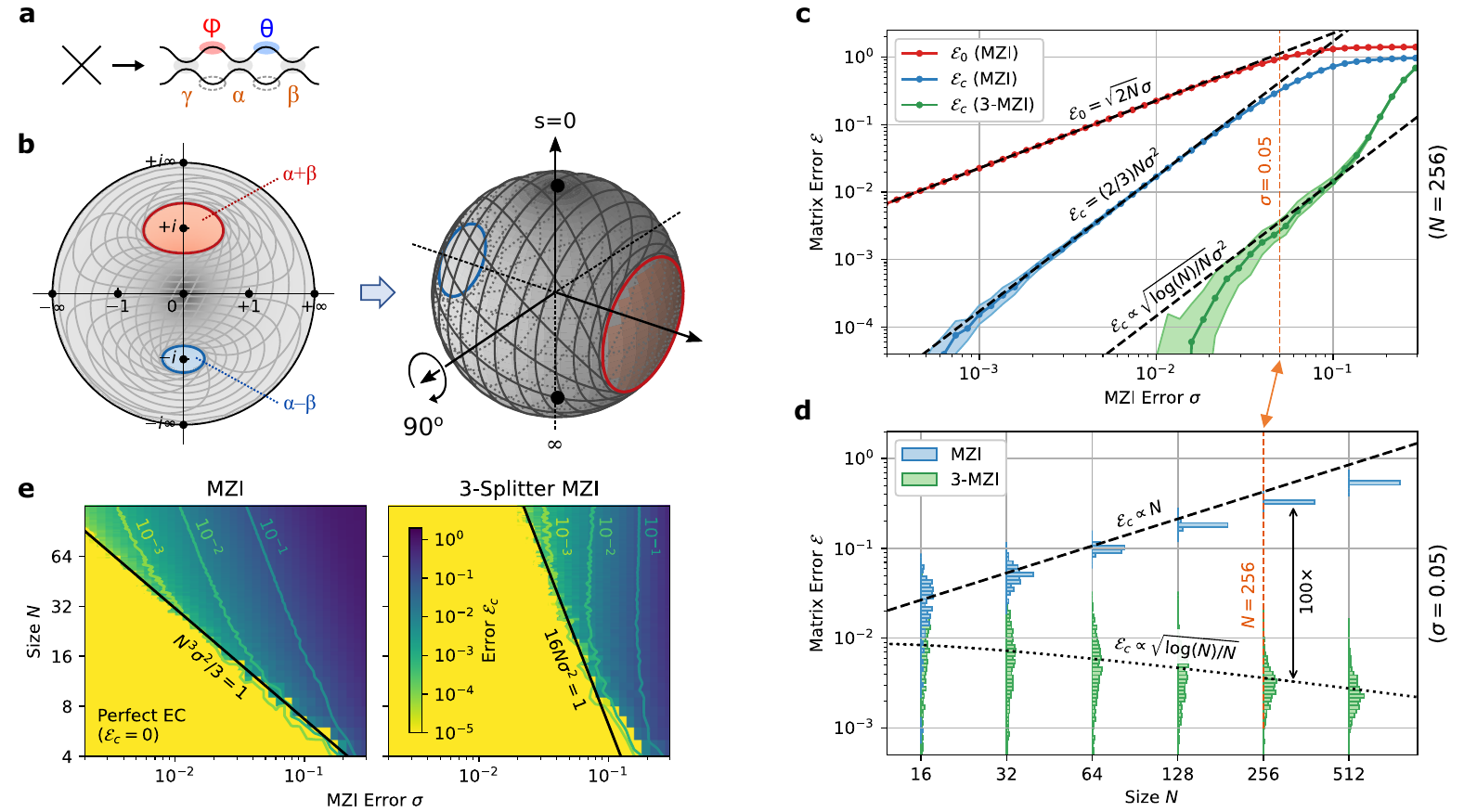}
\caption{3-splitter MZI design and simulated performance.  (a) Schematic of 3-MZI.  (b) Splitter M\"{o}bius transformation on $s \in \mathbb{C}$, which pushes the forbidden regions away from $s = \{0, \infty\}$, corresponding to a Riemann sphere rotation.  (c) Matrix error $\mathcal{E}_0$, $\mathcal{E}_c$ as a function of splitter variation $\sigma$ (fixed $N = 256$), comparing the standard and 3-splitter MZI designs.  (d) Scaling with mesh size $N$ (fixed $\sigma = 0.05$).  (e) Matrix error as function of both $N$ and $\sigma$, showing the sharp onset of ``perfect'' error correction in regions where the coverage $\mathcal{C}$ is of order unity.}
\label{fig:f4}
\end{center}
\end{figure*}

For linear photonic circuits, two important fidelity figures of merit are (1) the {\it coverage} $\mathcal{C}$, i.e.\ the probability that a matrix is realized exactly, and (2) the {\it normalized matrix error} $\mathcal{E} = \langle \lVert \Delta U \rVert_{\rm rms} \rangle/\sqrt{N}$, which is approximately equal to the average relative error for a given matrix element.  $\mathcal{C}$ and $\mathcal{E}$ depend on the error model and the distribution of target matrices.  Here, consistent with prior work \cite{Russell2017, Bandyopadhyay2021, RyanPaper1, RyanPaper2}, we sample target matrices randomly over the Haar measure \cite{Haar1933, Tung1985} and consider an uncorrelated Gaussian error model $\langle \alpha\rangle_{\rm rms} = \langle \beta\rangle_{\rm rms} = \sigma$.
Analytic expressions for $\mathcal{E}$ and $\mathcal{C}$ are derived in the Methods, which we summarize here.  If a mesh is straightforwardly programmed without taking any account of the imperfections (``uncorrected'' error), the normalized error is $\mathcal{E}_0 = \sqrt{2N}\sigma$ \cite{Bandyopadhyay2021, RyanPaper1}.  The coverage $\mathcal{C} = e^{-N^3\sigma^2/3}$ (Eq.~\ref{eq:cmzi}) decreases sufficiently fast that even moderately sized meshes have vanishingly small coverage, and error correction is generally imperfect.  In this case, the residual ``corrected'' error $\mathcal{E}_c = (2/3)N \sigma^2$ (Eq.~\ref{eq:ec0}) is the more relevant metric.  Since $\mathcal{E}_c \propto (\mathcal{E}_0)^2$, self-configuration correction affords a quadratic suppression of errors, which is a significant advantage when errors are below a threshold.  However, for sufficiently large meshes $N \gtrsim 1/\sigma^{2}$, error correction will be ineffective and the mesh cannot realize most matrices at high fidelity.  Thus, even with error correction, hardware imperfections set a fundamental scaling limit for standard MZI meshes.

\subsection*{Asymptotically Perfect Photonic Circuits}

%
The main challenge limiting error correction here is that the forbidden regions overlap with the peak of the probability distribution, which clusters tightly around the cross state $s = 0$ (Fig.~\ref{fig:f2}(e)) \cite{Russell2017}.  This clustering happens because light must propagate all the way down a mesh's diagonals to realize generic unitaries; the forbidden regions disrupt this ballistic transport leading to clipping of off-diagonal matrix elements \cite{Pai2019}.  Adding redundant components (MZI doubling) solves this problem by eliminating the forbidden regions altogether \cite{Miller2015, Suzuki2015}, but at the cost of added optical and electrical complexity.  Here, we take the alternative approach of {\it displacing} the forbidden regions away from the cross state.  This can be performed by placing a third splitter at the input of the MZI, as shown in Fig.~\ref{fig:f4}(a).  The extra splitter performs a M\"obius transformation $s \rightarrow (s + i \tan \eta)/(1 + i s \tan \eta)$, which for a 50:50 splitting ratio ($\eta = \pi/4$) maps the bar and cross states to $s = \pm i$ (Fig.~\ref{fig:f4}(b)).  This can be visualized as a 90$^{\rm o}$ rotation on the Riemann sphere, which pushes the forbidden regions to the equator, while the probability density is still concentrated at the poles (small errors $\gamma$ in the third splitter perturb this rotation angle slightly, but this does not change the structure of the forbidden regions and has little effect on the error correction).

This ``3-splitter MZI'' (3-MZI) can realize the full range of (absolute value) splitting ratios 
 $|s| \in [0, \infty)$, 
and can thus function as a high-contrast optical switch~\cite{Suzuki2015, Wang2020}.  However, the presence of forbidden regions means that the relative {\it phase} of this splitter cannot be fully controlled; which means that errors can still occur when programming the mesh (unlike the ``perfect'' MZIs of Refs.~\cite{Miller2015, Suzuki2015, Wilkes2016}, which cure this defect with redundant phase shifters).  However, from the distributions in Fig.~\ref{fig:f2}(e), for large meshes $\hat{s}$ will fall into the 3-MZI's forbidden regions only rarely.  The normalized matrix error, calculated in the Methods (Eq.~\ref{eq:ec3mzi}), takes the following form:
\beq
	\mathcal{E}_c \approx 8\sigma^2 \Bigl[2\frac{\log(N)-1.366}{N}\Bigr]^{1/2} \label{eq:ec3mzi2}
\eeq
In Fig.~\ref{fig:f4}(c-d), we numerically simulate self-configuration on imperfect meshes using the \textsc{Meshes} package (see Methods and supplemental code); the realized $\mathcal{E}_c$ shows good agreement with Eq.~(\ref{eq:ec3mzi2}).  For most mesh sizes, the $\mathcal{E}_c$ is 1--2 orders of magnitude smaller for the 3-MZI design.  Remarkably, the error actually {\it decreases} with increasing mesh size, scaling as $\mathcal{E}_c \propto \sqrt{\log(N)/N}$.  In the asymptotic limit $N \rightarrow \infty$, matrices can be programmed perfectly.


This non-intuitive effect arises from the fact that, under the Haar measure, only a small fraction of MZIs have significant probability density near $s = \pm i$, where the forbidden regions are centered \cite{Russell2017}.  
This probability   
decreases exponentially with the distance from the triangle's base (see Methods for details).  Therefore, although the mesh has $N(N-1)/2$ MZIs, only $O(N)$ contribute significantly to the matrix error under self-configuration.  A na\"{i}ve estimate assuming uncorrelated errors would give $\lVert \Delta U \rVert \propto \sqrt{N} \sigma^2$, which would lead to a constant $\mathcal{E}_c$.  However, during the self-configuration process, subsequent MZIs can partially correct for errors in earlier MZIs that cannot be properly configured; the end result is to reduce the overall error of each MZI by a factor proportional to $\sqrt{\log(N)/N}$ (see Methods), yielding the result Eq.~(\ref{eq:ec3mzi2}).

\begin{figure}[tb]
\begin{center}
\includegraphics[width=1.00\columnwidth]{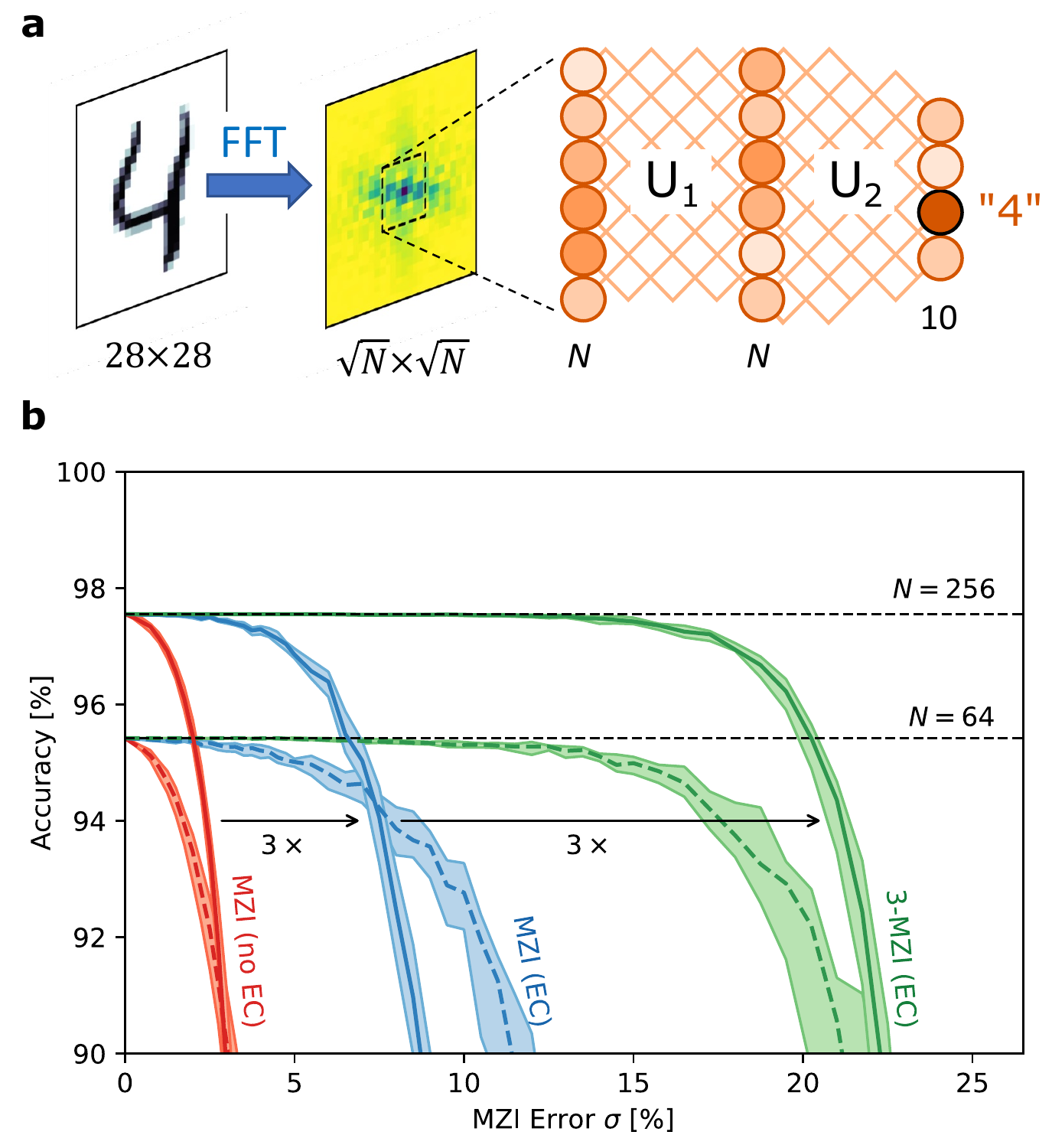}
\caption{Effect of hardware errors on DNN inference.  (a) Benchmark neural network consisting of FFT preprocessing, windowing, and two DNN layers, where the linear connections $U_1$ and $U_2$ are realized with MZI meshes \cite{RyanPaper2, Williamson2019}.  (b) Inference accuracy as a function of MZI error.}
\label{fig:f11}
\end{center}
\end{figure}

Another advantage of the 3-splitter MZI is that the threshold for perfect error correction is higher.  One obtains this threshold is found by computing the coverage $\mathcal{C} = e^{-16 N \sigma^2}$ (see Methods, Eq.~(\ref{eq:cov3})).  This is much larger than the coverage of the regular MZI mesh, and the threshold scales as $\sigma_{\rm th} \propto N^{-1/2}$, as opposed to the $N^{-3/2}$ scaling observed for the standard mesh.  Consequently, perfect error correction is available under a much wider range of conditions, as shown in Fig.~\ref{fig:f4}(e).

\subsection*{Error-Resilient Optical Neural Networks}

To highlight the significance of this error reduction, consider as a concrete example deep neural network (DNN) inference on coherent optical hardware.  A DNN is a sequence of layers, consisting of linear synaptic connections and nonlinear neuron activations.  An emerging application of photonics seeks to use optical interference to accelerate this process, encoding neuron activations in coherent optical amplitudes, while a programmable MZI mesh implements the synaptic weights and activations are performed with an all-optical or electo-optic nonlinearity \cite{Shen2017}.  Scaling remains the major challenge to constructing practical optical neural networks, as large mesh sizes ($N > 100$) are required to achieve a significant advantages over electronic hardware, and such large meshes are especially susceptible to fabrication errors.  A recent numerical study showed that even with state-of-the-art process tolerances, hardware errors can significantly degrade DNN inference accuracy \cite{Fang2019}, a difficulty that has spurred investigations into alternatives to the MZI mesh, which all have their own limitations \cite{Tait2017, Hamerly2019, Bernstein2022, Chen2022}.

Fig.~\ref{fig:f11}(a) depicts a benchmark neural network.  Here, $28\times 28$ images from the MNIST digit dataset \cite{LeCun1998} are preprocessed by a Fourier transform and cropped to a window of size $\sqrt{N}\times \sqrt{N}$, which forms the input to a two-layer unitary DNN.  The DNN can be implemented optically with rectangular MZI meshes for synaptic weighting \cite{Clements2016} and electro-optic nonlinearities for the activation (see Refs.~\cite{RyanPaper2, Williamson2019} for details).  Models with inner-layer sizes $N = 64$ and $N = 256$ are pre-trained using the \textsc{Neurophox} package \cite{Neurophox}, and inference accuracy is subsequently simulated on imperfect meshes with Gaussian splitter errors to calculate the classification accuracy.

This accuracy is plotted in Fig.~\ref{fig:f11}(b) for three cases: straightforwardly programming an MZI mesh without error correction, with error correction, and with the modified 3-MZI architecture.  Even for small device errors $\sigma = 1$--$2\%$, which is considered state-of-the-art for directional couplers in highly controlled fabrication processes \cite{Mikkelsen2014}, hardware errors significantly degrade the model's inference accuracy relative to its canonical value ($\sigma = 0$).  For small $\sigma$, this is recovered using error correction \cite{Bandyopadhyay2021, RyanPaper2}.  However, many broadband coupler designs \cite{Soldano1995, Maese2013, Wang2016, Ye2020, Morino2014, Lu2015} trade bandwidth for fabrication sensitivity and are in practice very sensitive to process variations, meaning larger splitter errors $\sigma \gtrsim 5\%$ are common.  In this moderate-error regime, error correction alone is not sufficient and the network shows reduced accuracy, a problem that becomes more pronounced as the size $N$ increases.  Moving to the 3-MZI architecture overcomes this limitation, enabling effectively error-free inference (relative to the canonical model) even out to very large splitter errors $\sigma \approx 10$--$15\%$, far beyond what is likely to be encountered in practice.

\subsection*{Broadband Mesh for Correlated Errors}

\begin{figure}[b!]
\begin{center}
\includegraphics[width=1.00\columnwidth]{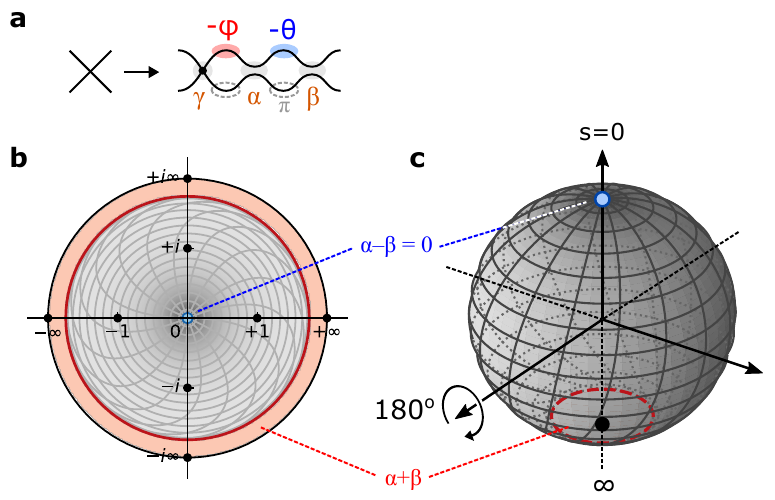}
\caption{MZI+Crossing architecture.  (a) Schematic of MZI+X.  (b) Effect of the crossing is to flip the $s = 0$ and $s = \infty$ forbidden regions.  For correlated errors, the forbidden region around $s = 0$ disappears.  (c) Riemann sphere projection.}
\label{fig:f7}
\end{center}
\end{figure}

%
For generic, uncorrelated component errors the 3-splitter MZI is well-suited.  However, since the correlation lengths of process variations tend to be larger than a single MZI \cite{Bogaerts2019}, errors are correlated in practice.  This is especially true for broadband couplers based on multimode interference (MMI) \cite{Soldano1995, Maese2013}, subwavelength gratings \cite{Wang2016, Ye2020}, and asymmetric designs \cite{Morino2014, Lu2015}, all of which are highly dependent on the device geometry, which can vary slightly from run to run.  Moreover, even with perfect 50:50 couplers, the splitting ratios are still wavelength-dependent.  Operating the mesh away from its design wavelength leads to correlated device errors, so sensitivity to these errors is closely tied to the operational bandwidth of the device.

Consider the case of a constant offset $\mu$ for all splitting ratios: $\alpha = \beta = \mu$.  In a standard MZI, the bar-state forbidden region (around $s = \infty$) disappears since $|\alpha - \beta| = 0$, while the cross-state region (around $s = 0$, the peak of the probability distribution) remains in place (Fig.~\ref{fig:f2}).  This is consistent with the common observation that the extinction ratio in an MZI is much higher in the cross port than in the bar port.  The optimal error reduction strategy, illustrated in Fig.~\ref{fig:f7}(a), was previously proposed in the context of broadband optical switching: place a waveguide {\it crossing} before the MZI  \cite{Suzuki2018}.  The added crossing performs the M\"{o}bius transformation $s \rightarrow 1/s$, rotating the Riemann sphere by 180$^{\rm o}$ to move the forbidden region to the minimum of the probability distribution (Fig.~\ref{fig:f7}(b-c)).

\begin{figure}[b!]
\begin{center}
\includegraphics[width=1.00\columnwidth]{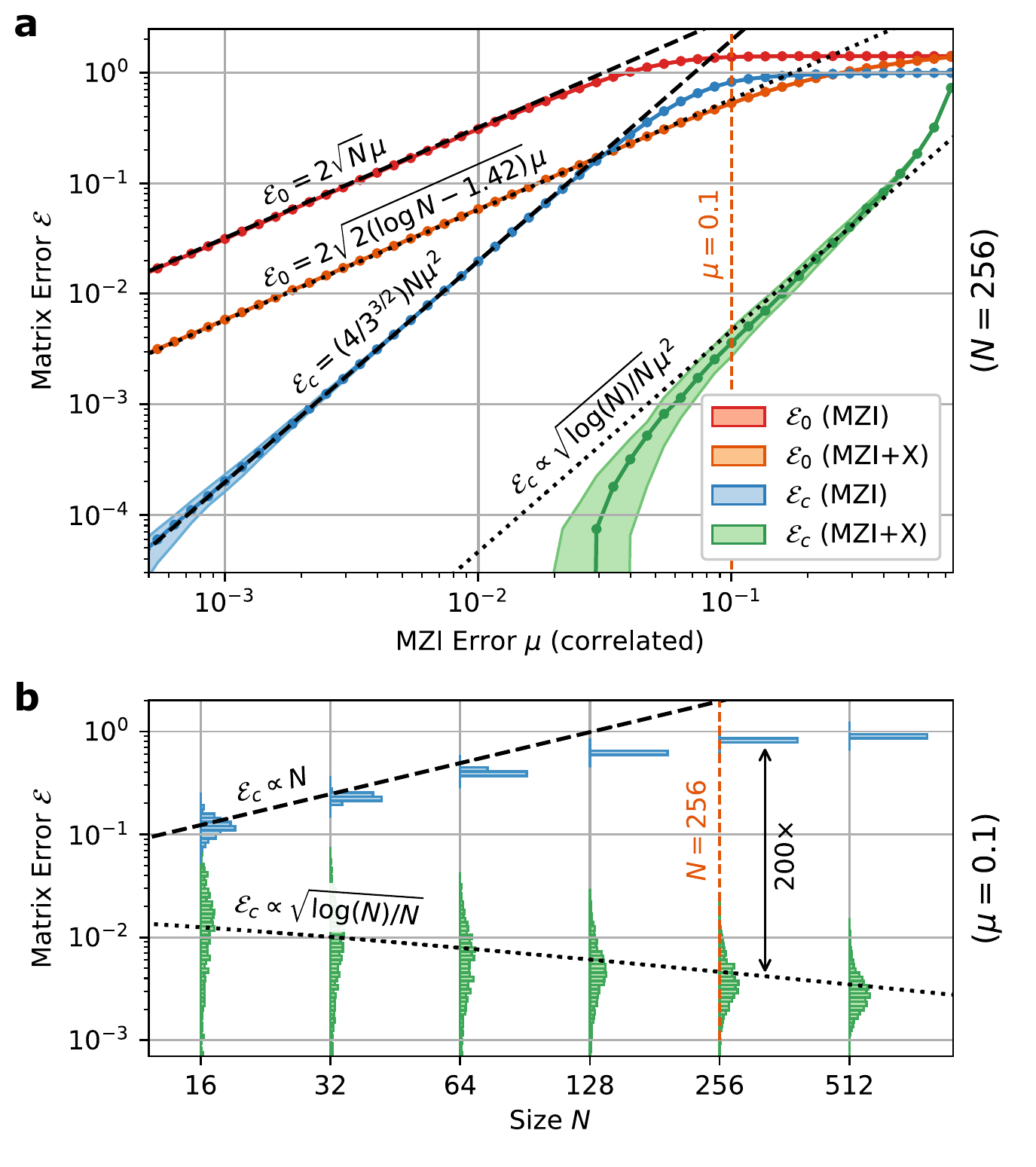}
\caption{Advantages of MZI+Crossing architecture for correlated component errors.  (a) Matrix error as a function of $\mu$ for fixed $N = 256$.  (b) Dependence on size $N$ for fixed $\mu$.}
\label{fig:f8}
\end{center}
\end{figure}

As before, we can calculate the coverage and matrix error of this ``MZI+Crossing'' (MZI+X) mesh by performing the nulling procedure on target unitaries, obtaining $\mathcal{C}$ from the probabilities that splitting ratios fall within the forbidden regions, and $\mathcal{E}_c$ from the residuals arising from imperfect diagonalization.  In this case, there is only one forbidden region, centered at $s = \infty$.  The calculation is worked out in the Methods.  For the normalized error, we find (Eq.~\ref{eq:ecmzix}):
\beq
	\mathcal{E}_c = 4\mu^2 \Bigl[\frac{2}{3} \frac{\log(N) - 0.423}{N} \Bigr]^{1/2} \label{eq:iv}
\eeq
This is plotted in Fig.~\ref{fig:f8}.  Like the 3-MZI design, this metric scales as $\mathcal{E}_c \propto \sqrt{\log(N)/N} \mu^2$, in contrast to the trend $\mathcal{E}_c = (4/3^{3/2}) N\mu^2$ calculated for the standard MZI under correlated errors.  The coverage also increases (Eq.~(\ref{eq:covmzix})), so that the threshold for perfect correction likewise scales as $\mu_{\rm th} \propto N^{-1/2}$, as opposed to  $\mu_{\rm th} \propto N^{-3/2}$ for the standard mesh.

Ultimately, the scalability of the MZI+X architecture is limited by differential errors $|\alpha - \beta|$ that arise from local fluctuations in waveguide dimensions.  The effect of such errors is analyzed in Supp.~Sec.~\SuppII.  For typical photonic process variations, $|\alpha - \beta| \ll \mu$ and differential errors are insignificant for mesh sizes up to at least $N = 512$.

\begin{table}[b!]
\begin{center}
\begin{tabular}{r|rrrrrr}
\hline\hline 
$N = $ & 16 & 32 & 64 & 128 & 256 & 512 \\
\hline
$F_{\rm TR} =$ & $5.6\times$ & $10\times$ & $18\times$ & $33\times$ & $61\times$ & $114\times$ \\
$F_{\rm BW} =$ & $2.4\times$ & $2.8\times$ & $3.4\times$ & $4.3\times$ & $5.6\times$ & $7.3\times$ \\
\hline\hline
\end{tabular}
\caption{Approximate tuning range and bandwidth enhancement factors 
for mesh sizes up to $N = 512$, Eqs.~(\ref{eq:fscal}, \ref{eq:tr}-\ref{eq:bw}).}
\label{tab:t1}
\end{center}
\end{table}

\begin{figure}[t]
\begin{center}
\includegraphics[width=1.00\columnwidth]{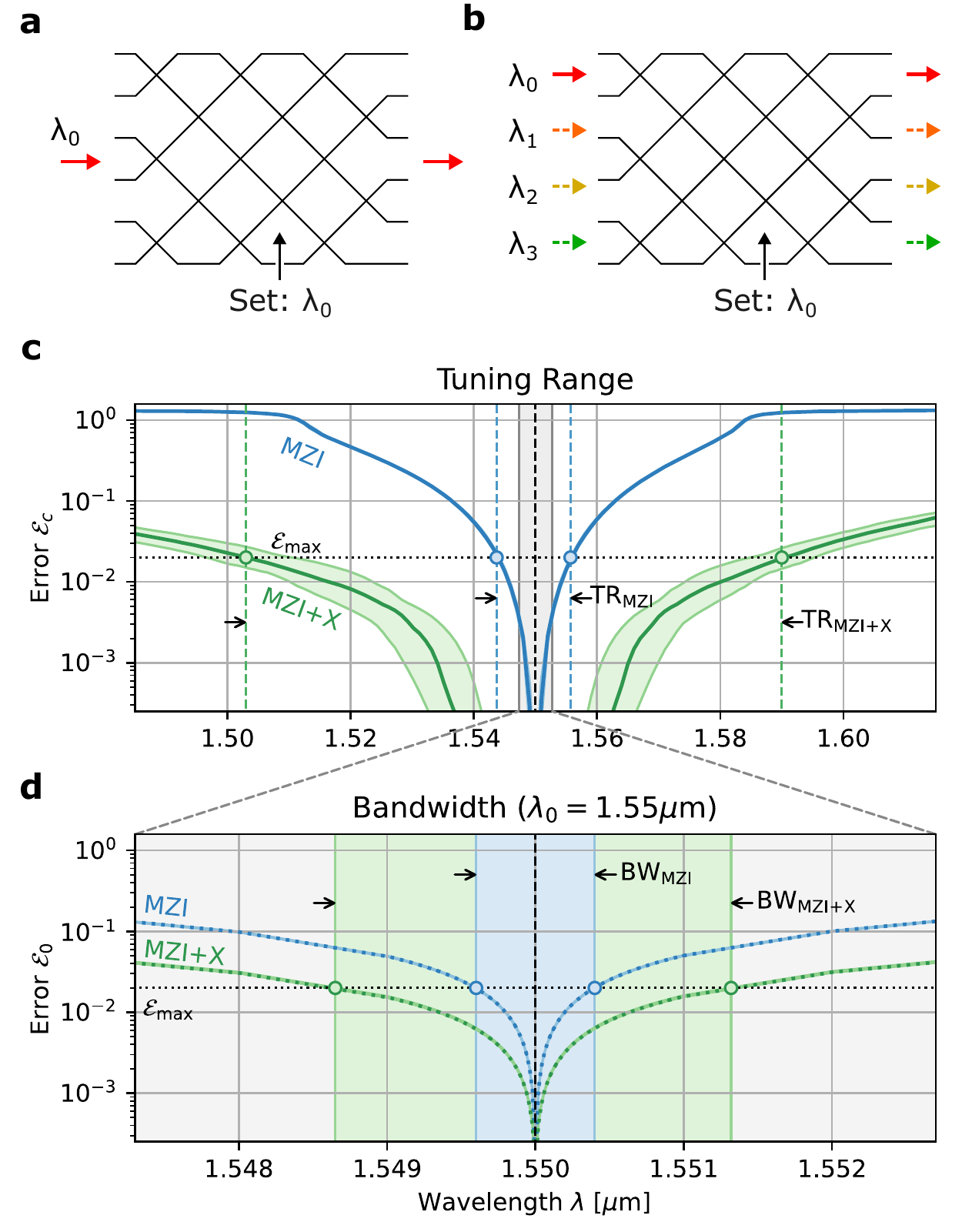}
\caption{Tuning range and bandwidth for MZI+X and standard MZI mesh, $N = 64$.  (a-b) Contrast between single- and multi-wavelength operation, which are limited by tuning range and bandwidth, respectively.  (c) Plot of $\mathcal{E}_c(\lambda)$, which dictates the tuning range for a target matrix error $\mathcal{E}_{\rm max}$.  (d) Corresponding plot of $\mathcal{E}_0(\lambda)$, which dictates the bandwidth.  Platform: $500\times 220$~nm Si:SiO$_2$ directional coupler with 200~nm gap, $\d\mu/d\lambda \approx 3.27/\mu\text{m}$.}
\label{fig:f9}
\end{center}
\end{figure}

As an added bonus, the MZI+X design also reduces the effect of errors in the absence of correction.  To see how, we can make an analogy to Bloch-sphere rotations.  The transfer matrix of a standard MZI is (up to a phase factor) the product of four rotations:
\beq
	T(\theta, \phi) \propto R_x(\tfrac\pi4+\mu) R_z(\theta) R_x(\tfrac\pi4+\mu) R_z(\phi)
\eeq
where $R_k(\eta) = e^{i \sigma_k \eta}$ is a Pauli rotation and $\sigma_k$ is a Pauli matrix.  For the cross state ($\theta = 0$), the errors $\mu$ add up constructively, while for the bar state ($\theta = \pi$), they cancel out (the latter is a simple example of dynamical decoupling of spins using a pulse sequence).  Most crossings in large meshes are close to the cross state, which leads to constructive addition of the errors in the standard MZI mesh.  However, for the MZI+X, the input ports of each MZI are exchanged, so the physical MZIs are close to the bar state where the errors cancel out.  The resulting uncorrected matrix error is (see Methods):
\beq
	\mathcal{E}_0 = \begin{dcases} 
		2\sqrt{N} \mu & \text{(MZI)} \\
		2\sqrt{2(\log N - 1.423)} \mu & \text{(MZI+X)}
		\end{dcases}
\eeq
Correlated errors (both corrected and uncorrected) are important because they are tightly connected to the operational bandwidth of the mesh, a critical design parameter for machine learning schemes that require broadband operation, e.g.\ for parallel processing on wavelength-multiplexed data \cite{Feldmann2021, Xu2021, Sludds2022, Davis2022}.  All beamsplitters are dispersive, and this dispersion leads to a correlated wavelength-dependent splitter error, which can usually be expanded to first order $\mu \approx (\d\mu/\d\lambda) \Delta\lambda$.  Two important wavelength-dependent figures of merit are (1) the {\it tuning range}, which refers to the range of $\lambda$ over which the mesh can be programmed to a given accuracy, Fig.~\ref{fig:f9}(a, c), and (2) the {\it bandwidth}, which is related to the number of wavelength channels that can be (simultaneously) processed by the mesh, Fig.~\ref{fig:f9}(b, d).  The tuning range is limited by the corrected error $\mathcal{E}_c$, while the bandwidth is limited by the uncorrected error $\mathcal{E}_0$, since a mesh cannot simultaneously error-correct at two different wavelengths.  Since the MZI+X design reduces both $\mathcal{E}_0$ and $\mathcal{E}_c$, it leads to enhancements in both the bandwidth and tuning range.  The enhancement factors scale as
\beq
	F_{\rm BW} \propto \sqrt{N/\log N},\ \ \ 
	F_{\rm TR} \propto (N^3/\log N)^{1/4} \label{eq:fscal}
\eeq
and are listed for several mesh sizes in Table~\ref{tab:t1} (see Methods for details).  As Fig.~\ref{fig:f9}(c-d) illustrates, the MZI+X architecture enjoys a significantly larger tuning range, in addition to modestly greater bandwidth.

Real crossings have a small amount of nonzero crosstalk, quantified by the S-matrix element $S_{21}$; scattering into the forward-facing port leads to a perturbation $R_x(\tfrac\pi2) \rightarrow R_x(\tfrac\pi2+\gamma)$ in the transfer matrix, where $\gamma = 10^{-S_{21}[\text{dB}]/20}$.  This does not degrade the effectiveness of self-configuration, since the additional scattering angle merely rotates the Riemann sphere Fig.~\ref{fig:f7}(c) by an additional angle $\gamma \ll 1$, and the forbidden region is still far from $s = 0$.  In-plane crossings in silicon can achieve sub-40~dB crosstalk suppression ($\gamma < 0.01$) with insertion losses well below 0.1~dB \cite{Fukazawa2004, Chen2006, Ma2013, Dumais2017, Wu2020}.  Unlike directional couplers, crossings are inherently broadband; the insertion loss and crosstalk depend only very weakly on $\lambda$, so any crossing imperfections can be treated as (correctable) wavelength-independent errors that do not affect the bandwidth enhancements of the MZI+Crossing scheme.  In addition to the forward-scattered light, a 90$^{\rm o}$ crossing will scatter light into the backward-facing port.  Back-reflected light can be subsequently reflected in other crossings, leading to a spurious signal that interferes with the forward-propagating light. Provided that the phases of reflected beams are random, these add in quadrature: with amplitude $\gamma^2$ and $O(N^2)$  scattering paths, we expect this to induce an $O(N\gamma^2)$ error, which may be uncorrectable and set a limit on scaling.  However, if this effect is small, gradient-based methods or iterative self-configuration may enable correction of these errors.

\section*{Discussion}
\begin{figure}[tbp]
\begin{center}
\includegraphics[width=1.00\columnwidth]{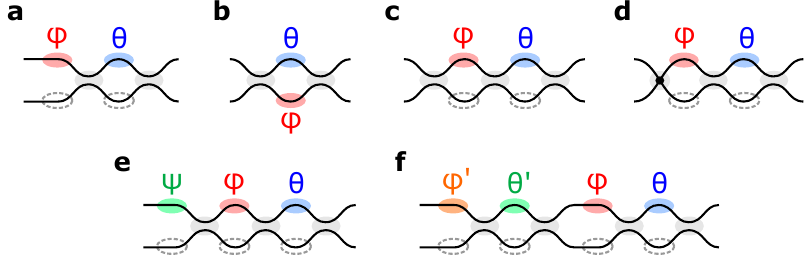}
\caption{Comparison of crossing types.  (a) MZI, (b) Symmetric (S-MZI) \cite{Bell2021}, (c) 3-splitter (3-MZI) \cite{Wang2020}, (d) Port-exchanged (MZI+X) \cite{Suzuki2018}, (e) Suzuki \cite{Suzuki2015}, (f) Miller \cite{Miller2015}.}
\label{fig:f10}
\end{center}
\end{figure}

%
As photonic circuits grow larger, error tolerance becomes increasingly important.  Many techniques exist to manage hardware errors, but all involve a tradeoff between accuracy and complexity.  At opposite poles lie ``zero-change'' error correction, which has limited scalability \cite{RyanPaper1, RyanPaper2, Bandyopadhyay2021, SriPaper}, and ``perfect'' photonic circuits, which require a larger number of photonic and electronic components \cite{Miller2015, Suzuki2015}.  This paper has introduced two designs for programmable circuits that strike a tradeoff between these extremes, as shown in Fig.~\ref{fig:f10} and Table~\ref{tab:t2}, achieving performance that is almost as good as the perfect designs, but with less added complexity (see Supp.~Sec.~\SuppI~for details).

The main insight from this paper is that, by adding a single passive component (either a splitter or a waveguide crossing) to the MZI, we can recover behavior that is asymptotically perfect---that is, the average normalized matrix error {\it decreases} with size.  Our design choices are motivated by the elegant theory of self-configuration by matrix diagonalization \cite{RyanPaper2}, where splitting ratios are set to successively zero the off-diagonal elements of the target unitary.  By visualizing the MZI state on the Riemann sphere, we can intuitively understand the increased error robustness of our designs in terms of ``rotating'' the forbidden regions away from the peak probability density.  This leads to a several-orders-of-magnitude reduction in post-correction errors compared to the standard MZI mesh.  The ability to achieve near-perfect and freely scalable MZI meshes with less complexity than the MZI-doubled designs \cite{Miller2015, Suzuki2015} (especially with respect to the number of active components and pads) removes a major obstacle to the realization of very-large-scale photonic circuits.

An interesting direction for future work is to explore to what extent multiport interferometers can be made robust to imperfections in the absence of error correction.  For example, previous studies of 3-MZI splitters have noted a wavelength-independent coupling ratio for certain parameter choices \cite{Wang2020}.  Likewise, the near-cancellation of correlated errors in the MZI+Crossing architecture explains the $O(\sqrt{N/\log N})$ reduction in the uncorrected error, and corresponding increase in bandwidth.  Further design modifications based on the theory of composite pulse sequences \cite{Brown2004, Bulmer2020, Little1997} may allow this imperfect cancellation to be made exact, further improving the bandwidth (and multiplexing capabilities) of linear photonics.

\begin{table}[tbp]
\begin{center}
\begin{tabular}{c|ccc|ccc}
\hline\hline
& \multicolumn{3}{c|}{\bf Complexity} & \multicolumn{3}{c}{\bf Features$^\dagger$} \\
& Passives & Actives & Area & & & \\ \hline
MZI & 2 & 2 & 1.0 & S\! & \\
S-MZI & 2 & 2 & 0.8 & & \\ \hline
3-MZI & 3 & 2 & 1.2 & S\! & & \!\!\!(P)\! \\
MZI+X & 3 & 2 & 1.2 & S\! & \!B\! & \!\!\!(P)\! \\ \hline
Suzuki & 3 & 3 & 1.5 & S\! & & \!\!P \\
Miller & 4 & 4 & 2.0 & S\! & & \!\!P \\ \hline\hline
\end{tabular}
\caption{Characteristics of the major tunable crossing types.  $^\dagger$S: Self-configuration.  B: Broadband.  (P): Asymptotically perfect.  P: Perfect.}
\label{tab:t2}
\end{center}
\end{table}

\small

\section*{Methods}

\subsection*{Unitaries and the Riemann Sphere}

A generic $2\times 2$ complex-valued matrix has eight degrees of freedom, and a $2\times 2$ unitary has four.  However, the space of $2\times 2$ unitaries can be divided into equivalence classes based on the splitting ratio $s = T_{11}/T_{12}$.  Specifically, any two unitaries are equivalent up to output phases, i.e.\ $T = \text{diag}(e^{i\psi_1}, e^{i\psi_2}) \hat{T}$, if and only if the splitting ratios are the same, $s = \hat{s}$.  As a complex number, $s$ can be visualized on the Riemann sphere (Fig.~\ref{fig:f2}(d)), where the mapping is performed by the stereographic projection $s = (x + iy)/(1 + z)$ (which inverts to $x + iy = 2s/(1+|s|^2)$, $z = (1+|s|^2)/(1-|s|^2)$).

Ordinarily, the distance between matrices is defined as the Frobenius ($L_2$) norm $\lVert \Delta U \rVert = (\sum_{mn}|\Delta U_{mn}|^2)^{1/2}$.  However, since output phases are corrected in subsequent steps, the most relevant distance metric for a $2\times 2$ block is the Frobenius norm modulo these phase shifts,
\beq
	d(T, \hat{T}) \equiv \text{min}_\psi\Bigl\lVert T - \begin{bmatrix} e^{i\psi_1}\!\! & \\ & \!\! e^{i\psi_2} \end{bmatrix} \hat{T} \Bigr\rVert
	= \frac{d(s, \hat{s})}{\sqrt{2}} \label{eq:dtt}
\eeq
where $d(s, \hat{s}) = 2|s - \hat{s}|/\sqrt{(|s|^2+1)(|\hat{s}|^2+1)}$ is the Euclidean distance between two points on the Riemann sphere.

A common parameterization is $s = e^{i\phi} \tan(\theta/2)$, which represents the splitting ratio of the standard MZI, Fig.~\ref{fig:f1}(b).  On the Riemann sphere, $(\theta, \phi)$ map to the standard polar coordinates, i.e.\ $x = \sin(\theta)\cos(\phi)$, $y = \sin(\theta)\sin(\phi)$, $z = \cos(\theta)$.


\subsection*{Coverage and Matrix Error Derivation}

The nulling method relies on successive zeroing of off-diagonal elements to diagonalize the matrix $X$ (initialized to $U$).  Each nulling step zeros a single element, increasing the size of the zeroed-out off-diagonal region.  Nulling steps are performed in a particular order to ensure that zeroed-out elements remain zero after all subsequent steps \cite{RyanPaper2, Reck1994, Clements2016}.  In a given step, if nulling cannot be achieved perfectly, the ``zeroed-out'' region of matrix $X$ is left with a residual of magnitude:
\beq
	r = |T_{11}v - T_{12}u| = \sqrt{|u|^2 + |v|^2} \frac{d(s, \hat{s})}{2} \label{eq:res}
\eeq
where $\hat{s}$ is the target splitting ratio, $s$ is the closest physically realizable value, and $d(s, \hat{s})$ is the Euclidean distance on the Riemann sphere, the same metric used in Eq.~(\ref{eq:dtt}).

The coverage and matrix error depend on (1) the distribution $P(s)$ of target splitting ratios, a function of the distribution of target unitaries, and (2) the locations and sizes of the forbidden regions, a function of the specific mesh implementation (MZI, 3-MZI, MZI+X).  For the Haar measure, $P(s)$ depends on an MZI's location in the mesh; for a given $T_{mn}$ it takes the following form \cite{Russell2017}:
\beq
	P_{mn}(s) = \frac{n}{4\pi} \Bigl(\frac{z+1}{2}\Bigr)^{n-1}
	= \frac{n}{4\pi(1+|s|^2)^{n-1}}
	 \label{eq:pmn}
\eeq
Here, the density is defined with respect to the area measure on the Riemann sphere
\beq
	d\mu = \sin(\theta) \d\theta\d\phi = \frac{4}{(1+|s|^2)} \d^2s
\eeq
so that $\int{P_{mn}(s) \d\mu(s) = 1}$.  Note that, under Eq.~(\ref{eq:pmn}), $P_{mn}$ is uniform for the lowest row of crossings, and becomes increasingly concentrated as one approaches the triangle's apex; as a result, the overall distribution is strongly biased towards the cross state for large meshes, as shown in Fig.~\ref{fig:f2}(e) (the same distribution also holds for the rectangular mesh, up to a reordering of the MZIs).

The forbidden regions $\mathcal{F}_\pm$ are centered at opposite poles of the Riemann sphere
\beq
	(s_+,\ s_-) = \begin{cases}
		(0,\ \infty) & \text{(MZI)} \\
		(+i,\ -i) & \text{(3-MZI)} \\
		(\infty,\ 0) & \text{(MZI+X)}
		\end{cases}
\eeq
and have radii $R_\pm = 2|\alpha \pm \beta|$.  In the case of small hardware errors, where $P(s) \approx P(s_\pm)$ inside each $\mathcal{F}_\pm$, the probability that $\hat{s}$ falls inside the region is given by $\pi R_\pm^2 P(s_\pm)$.  The coverage $\mathcal{C}$ is the probability that every $\hat{s}$ avoids the forbidden regions, and is well approximated by
\beq
	\mathcal{C} = \exp\Bigl(-\sum_{mn} \pi \bigl(P_{mn}(s_+) \langle R_+^2 \rangle + P_{mn}(s_-) \langle R_-^2 \rangle \bigr) \Bigr) \label{eq:cov0}
\eeq

The normalized matrix error $\mathcal{E}_c = \langle \lVert \Delta U \rVert_{\rm rms} \rangle/\sqrt{N}$ is approximately the quadrature sum of the residuals accumulated during nulling:
\beq
	(\mathcal{E}_c)^2 = \frac{\langle \lVert \Delta U \rVert^2 \rangle }{N} = \frac{2}{N} \sum_{mn} \langle r_{mn}^2 \rangle \label{eq:ec}
\eeq
Here, $\langle \ldots \rangle$ refers to the ensemble average over both Haar-distributed target unitaries $U$ \cite{Haar1933, Tung1985} and the distribution of hardware errors $\alpha, \beta$.  We calculate the mean residual $\langle r^2 \rangle$ by averaging Eq.~(\ref{eq:res}) over the distribution $P(s)$.  This is simplified in the case of small hardware errors, because the forbidden region is correspondingly small and where we can assume $P(s)$ is approximately constant:
\beq
	\langle r_{mn}^2 \rangle = 
	\frac{\pi}{24} \underbrace{\bigl\langle|u|^2 + |v|^2\bigr\rangle}_{q_{mn}} \bigl[P_{mn}(s_+) \langle R_+ \rangle^4 + P_{mn}(s_-) \langle R_- \rangle^4\bigr] \label{eq:res2}
\eeq
This residual depends on the quantity $q_{mn} = \langle |u|^2 + |v|^2 \rangle$, where $(u, v)$ are the highlighted in green in Fig.~\ref{fig:f2}(b).  Following the Gaussian elimination procedure of a Haar matrix, this evaluates to $q_{mn} = (n+1)/(N+1-m)$.

A detailed description of the nulling algorithm, including a comparison to the local method \cite{Bandyopadhyay2021} and global optimization \cite{Burgwal2017, Mower2015, Pai2019} (which has a much longer convergence time), is presented in Supp.~Sec.~\SuppIII.

\subsection*{Gaussian Errors: MZI \& 3-MZI}

For the uncorrelated Gaussian perturbation model with $\langle \alpha \rangle_{\rm rms} = \langle \beta \rangle_{\rm rms} = \sigma$, the forbidden regions are (statistically) symmetric, with moments $\langle R_\pm^2 \rangle = 8\sigma^2$ and $\langle R_\pm^4 \rangle = 192\sigma^4$.  

For the MZI mesh, the coverage expression Eq.~(\ref{eq:cov0}) is dominated by the $s = 0$ term, where $P_{mn}(0) = n/4\pi$.  Considering only this term, we calculate:
\beq
	\mathcal{C}_{\rm MZI} = \exp\Bigl(-\frac{\langle R_+^2 \rangle}{4} \sum_{mn} n\Bigr) \rightarrow e^{-N^3\sigma^2/3} \label{eq:cmzi}
\eeq
where we have replaced the discrete sum by an integral
\beq
	\sum_{mn} (\ldots) \rightarrow \int_0^N \int_0^{N-m} (\ldots) \d n\,\d m \label{eq:intsubs}
\eeq
which is valid in the limit of large $N$.  Likewise, the top forbidden region dominates the matrix error, so we evaluate Eq.~(\ref{eq:res2}) including only the first term in the sum:
\beq
	\langle r_{mn}^2 \rangle_{\rm MZI} \rightarrow \frac{n(n+1)}{N+1-m} \frac{\langle R_+^4 \rangle}{96}
\eeq
Converting the sum to an integral and substituting $\langle R_+^4 \rangle$, we find:
\beq
	(\mathcal{E}_{c})_{\rm MZI} = \sqrt{\frac{N^2}{432} \langle R_+^4 \rangle} \rightarrow \frac{2}{3} N\sigma^2 \label{eq:ec0}
\eeq
Now we redo the calculation for the 3-MZI.  In this case, the forbidden regions are located at $s_\pm = \pm i$ and contribute equally to the problem.  Following Eq.~(\ref{eq:cov0}), the coverage is given by:
\beq
	\mathcal{C}_{\text{3-MZI}} = \exp\Bigl(-2 \times \sum_{mn} \pi \langle R_\pm^2 \rangle P_{mn}(\pm i) \Bigr) \rightarrow e^{-16 N \sigma^2} \label{eq:cov3}
\eeq
Applying Eq.~(\ref{eq:res2}), the mean residual left by crossing $T_{mn}$ is:
\beq
	\langle r_{mn}^2\rangle_{\text{3-MZI}} = 2\times \frac{\pi}{24} \underbrace{\frac{n+1}{N+1-m}}_{q_{mn}} \underbrace{\frac{n}{2^{n+1}\pi}}_{P_{mn}(\pm i)} \underbrace{\vphantom{\frac{n}{n}}(192\sigma^4)}_{\langle R_+^4\rangle} \label{eq:res3}
\eeq
The factors of two in Eqs.~(\ref{eq:cov3}-\ref{eq:res3}) arise because both forbidden regions contribute equally.  This $\langle r_{mn}^2 \rangle$ is not slowly-varying with $(m, n)$, so we cannot convert the sums to integrals.  We first perform the summation over $n$, which converges rapidly due to the $1/2^{n+1}$ factor (approximating the upper bound to infinity because of the rapid convergence), followed by summation over $m$.  We find the normalized error:
\begin{align}
	(\mathcal{E}_c)_\text{3-MZI} & = \Bigl(\frac{128 \sigma^4}{N} \Bigl[\sum_{n=1}^N \frac{1}{n} - \frac{5}{4} - \log(2)\Bigr]\Bigr)^{1/2} \nonumber \\
	& \approx 8\sigma^2 \Bigl[2 \frac{\log(N) + \gamma_e - \tfrac{5}{4} - \log(2)}{N}\Bigr]^{1/2} \label{eq:ec3mzi}
\end{align}
where $\gamma_e \approx 0.5772$ is the Euler-Mascheroni constant.

\subsection*{Correlated Errors: MZI \& MZI+X}

Under a correlated error model, $\alpha = \beta = \mu$.  In this case, there is only one forbidden region, which for the MZI is centered at $s_+ = 0$, with $R_+ = 4\mu$.  The coverage and matrix error for the standard MZI can then be calculated from Eqs.~(\ref{eq:cmzi}, \ref{eq:ec0}) with the appropriate substitutions for $\langle R_+^2 \rangle$, $\langle R_+^4 \rangle$:
\begin{align}
	\mathcal{C}_{\rm MZI} & = e^{-(2/3) N^3\mu^2}  \\
	(\mathcal{E}_c)_{\rm MZI} & = (4/3^{3/2}) N\mu^2
\end{align}
Now consider the MZI+X.  The additional crossing rotates the forbidden region to $s_+ \rightarrow \infty$.  Only the MZIs in the bottom row of the triangle ($n = 1$) contribute to the sums in Eqs.~(\ref{eq:cov0}-\ref{eq:ec}), because the probability distribution Eq.~(\ref{eq:pmn}) vanishes at $s = \infty$ for the upper rows.  

As before, we use the residual formula Eq.~(\ref{eq:res2}) to calculate the matrix error.  In this case, there is only one forbidden region, centered at $s_+ = \infty$, with $R_+ = 4\mu$.  Only the MZIs in the bottom row contribute to the sum, because the probability distribution Eq.~(\ref{eq:pmn}) vanishes at $s = \infty$ for the upper rows.  The coverage is:
\beq
	\mathcal{C}_{\text{MZI+X}} = \exp\Bigl(-\sum_{m} \pi \langle R_+^2 \rangle P_{m1}(\infty) \Bigr) \rightarrow e^{-4N\mu^2} \label{eq:covmzix}
\eeq
With the mean residual given by
\beq
	\langle r_{m1}^2\rangle_{\text{MZI+X}} = \frac{\pi}{24} \underbrace{\frac{2}{N+1-m}}_{q_{m1}} \!\underbrace{\vphantom{\frac{2}{N+1-m}} \frac{1}{4\pi}}_{P_{m1}(\infty)} \!
	\underbrace{\vphantom{\frac{2}{N+1-m}}(256\mu^4)}_{\langle R_+^4\rangle} \label{eq:resX}
\eeq
and $\langle r_{mn}^2\rangle = 0$ for $n > 1$, the matrix error evaluates to:
\beq
	(\mathcal{E}_c)_\text{MZI+X} = 4\mu^2 \Bigl[\frac{2}{3} \frac{\log(N) + \gamma_e - 1}{N} \Bigr]^{1/2} \label{eq:ecmzix}
\eeq
Now we consider the uncorrected matrix error.  For the standard MZI mesh, this is $\mathcal{E}_0 = 2\sqrt{N}\mu$ \cite{RyanPaper2}.  Using the transfer matrix of the standard MZI
\beq
	T_{\alpha,\beta}(\theta,\phi) = R_x(\tfrac\pi4\!+\!\beta) \begin{bmatrix} e^{i\theta} & 0 \\ 0 & 1 \end{bmatrix} R_x(\tfrac\pi4\!+\!\alpha) \begin{bmatrix} e^{i\phi} & 0 \\ 0 & 1 \end{bmatrix}
\eeq
to first order in $(\alpha, \beta)$, the norm of the matrix error is:
\beq
	\lVert \Delta T \rVert_{\rm MZI}^2 = 2 \bigl[\cos^2(\theta/2) (\alpha+\beta) + \sin^2(\theta/2) (\alpha-\beta)^2\bigr]
\eeq
which is maximized when the MZI is in the cross state $\theta = 0$.  For the MZI+Crossing (Fig.~\ref{fig:f7}(a)), we find:
\begin{align}
	& T_{\alpha,\beta}^{\text{(X)}}(\theta,\phi) = R_x(\tfrac\pi4\!+\!\beta) \!\begin{bmatrix} e^{-i\theta}\! & 0 \\ 0 & \!-1 \end{bmatrix}\! R_x(\tfrac\pi4\!+\!\alpha) \!\begin{bmatrix} e^{-i\phi}\! & 0 \\ 0\! & 1 \end{bmatrix}\! R_x(\tfrac\pi2) \nonumber \\
	& \quad = e^{-i(\theta+\phi)} \begin{bmatrix} 1 & 0 \\ 0 & -1 \end{bmatrix} T_{\alpha,-\beta}(\theta,\phi)
\end{align}
Up to irrelevant output phases, the effect of the crossing is to flip the relative sign of $\alpha$ and $\beta$, so the component errors appear anticorrelated.  As a result, $\lVert \Delta T\rVert_{\text{MZI+X}} \propto \sin(\theta/2) \mu$, which is zero for the cross state.  The actual error is found by adding the $\lVert \Delta T_{mn} \rVert$ in quadrature and averaging over the probability distribution $P_{mn}(\theta) = n \sin(\theta/2)\cos(\theta/2)^{2n-1}$ (equivalent to Eq.~(\ref{eq:pmn})):
\beq
	\mathcal{E}_0 = 2\sqrt{2(\log N + \gamma_e - 2)} \mu \label{eq:e0mzix}
\eeq
For a wavelength-dependent splitter error $\mu \approx (\d\mu/\d\lambda) \Delta\lambda$, the tuning range and bandwidth can be calculated from the expressions for $\mathcal{E}_c$ (Eq.~(\ref{eq:ecmzix})) and $\mathcal{E}_0$ (Eq.~(\ref{eq:e0mzix})), respectively: the tuning range is the range over which $\mathcal{E}_c(\lambda) < \mathcal{E}_{\rm max}$, while the bandwidth is the range over which $\mathcal{E}_0(\lambda) < \mathcal{E}_{\rm max}$:
\begin{align}
	\Delta\lambda_{\rm TR} & = \frac{\sqrt{\mathcal{E}_{\rm max}}}{|\d\lambda/\d\mu|} 
			\begin{cases}
				\frac{3^{3/4}}{\sqrt{N}} & \text{(MZI)} \\
				\sqrt{\frac{3N}{2(\log N-0.42)}} & \text{(MZI+X)}
			\end{cases} \label{eq:tr} \\
	\Delta\lambda_{\rm BW} & = \frac{\mathcal{E}_{\rm max}}{|\d\lambda/\d\mu|} 
			\begin{cases}
				\frac{1}{\sqrt{N}} & \text{(MZI)} \\
				\frac{1}{\sqrt{2(\log N - 1.42)}} & \text{(MZI+X)}
			\end{cases} \label{eq:bw}
\end{align}
From these expressions, we derive the enhancement factors reported in Eqs.~(\ref{eq:fscal}) and Table~\ref{tab:t1}.

\subsection*{Neural Network Model}

The optical neural network model is based on the architecture described in Ref.~\cite{Pai2020}.  Input images are first Fourier transformed, and cropped to a $\sqrt{N}\times \sqrt{N}$ window, where $N$ is the DNN's inner layer size.  The signal from this window ($N$ input neurons) passes through two optical layers, with unitary connectivity realized with rectangular meshes.  The activation function at the inner layer is realized electro-optically: a fraction of each output field is tapped off and sent to a detector, whose photocurrent modulates the remaining output light \cite{Williamson2019, Bandyopadhyay2022}, implementing the activation function:
\beq
	f(E) = \sqrt{1-\alpha}\, e^{-i(g|E|^2 + \phi - \pi)/2} \cos\bigl(\tfrac{1}{2}(g|E|^2 + \phi)\bigr)
\eeq
where $\alpha$ is the power tap fraction, $g$ is the modulator response, and $\phi$ is the phase at zero power.  Here, we choose $\alpha = 0.1$, $g = \pi/20$, and $\phi = \pi$, so that $f(E)$ approximates a leaky ReLU in the right power regime.  Models of sizes $N = 64$ and $N = 256$ were trained using the \textsc{Neurophox} package \cite{Neurophox}.

\subsection*{Simulations and Data Analysis}

All simulations were performed using the \textsc{Meshes} package, an open-source simulator for feedforward photonic circuits that can account for hardware imperfections \cite{Meshes}.  Figs.~\ref{fig:f4}, \ref{fig:f11}, \ref{fig:f8}, \ref{fig:f9} plot multiple instances (usually $\geq 100$) per point; dots show medians while shaded regions show the interquartile range.  Source code to produce the plots for this manuscript is provided in the Supplementary Material.

\section*{Data Availability}

All data from this paper can be generated using the \textsc{Meshes} package \cite{Meshes} and source code files provided in the Supplementary Material.

\section*{Code Availability}   

Source code files are provided in the Supplementary Material.

\normalsize

%

\begin{thebibliography}{10}
\expandafter\ifx\csname url\endcsname\relax
  \def\url#1{\texttt{#1}}\fi
\expandafter\ifx\csname urlprefix\endcsname\relax\def\urlprefix{URL }\fi
\providecommand{\bibinfo}[2]{#2}
\providecommand{\eprint}[2][]{\url{#2}}

\bibitem{Reck1994}
\bibinfo{author}{Reck, M.}, \bibinfo{author}{Zeilinger, A.},
  \bibinfo{author}{Bernstein, H.~J.} \& \bibinfo{author}{Bertani, P.}
\newblock \bibinfo{title}{Experimental realization of any discrete unitary
  operator}.
\newblock \emph{\bibinfo{journal}{Physical Review Letters}}
  \textbf{\bibinfo{volume}{73}}, \bibinfo{pages}{58} (\bibinfo{year}{1994}).

\bibitem{Clements2016}
\bibinfo{author}{Clements, W.~R.}, \bibinfo{author}{Humphreys, P.~C.},
  \bibinfo{author}{Metcalf, B.~J.}, \bibinfo{author}{Kolthammer, W.~S.} \&
  \bibinfo{author}{Walmsley, I.~A.}
\newblock \bibinfo{title}{Optimal design for universal multiport
  interferometers}.
\newblock \emph{\bibinfo{journal}{Optica}} \textbf{\bibinfo{volume}{3}},
  \bibinfo{pages}{1460--1465} (\bibinfo{year}{2016}).

\bibitem{Carolan2015}
\bibinfo{author}{Carolan, J.} \emph{et~al.}
\newblock \bibinfo{title}{Universal linear optics}.
\newblock \emph{\bibinfo{journal}{Science}} \textbf{\bibinfo{volume}{349}},
  \bibinfo{pages}{711--716} (\bibinfo{year}{2015}).

\bibitem{Zhong2020}
\bibinfo{author}{Zhong, H.-S.} \emph{et~al.}
\newblock \bibinfo{title}{Quantum computational advantage using photons}.
\newblock \emph{\bibinfo{journal}{Science}}  (\bibinfo{year}{2020}).

\bibitem{Shen2017}
\bibinfo{author}{Shen, Y.} \emph{et~al.}
\newblock \bibinfo{title}{Deep learning with coherent nanophotonic circuits}.
\newblock \emph{\bibinfo{journal}{Nature Photonics}}
  \textbf{\bibinfo{volume}{11}}, \bibinfo{pages}{441} (\bibinfo{year}{2017}).

\bibitem{Marpaung2013}
\bibinfo{author}{Marpaung, D.} \emph{et~al.}
\newblock \bibinfo{title}{Integrated microwave photonics}.
\newblock \emph{\bibinfo{journal}{Laser \& Photonics Reviews}}
  \textbf{\bibinfo{volume}{7}}, \bibinfo{pages}{506--538}
  (\bibinfo{year}{2013}).

\bibitem{Zhuang2015}
\bibinfo{author}{Zhuang, L.}, \bibinfo{author}{Roeloffzen, C.~G.},
  \bibinfo{author}{Hoekman, M.}, \bibinfo{author}{Boller, K.-J.} \&
  \bibinfo{author}{Lowery, A.~J.}
\newblock \bibinfo{title}{Programmable photonic signal processor chip for
  radiofrequency applications}.
\newblock \emph{\bibinfo{journal}{Optica}} \textbf{\bibinfo{volume}{2}},
  \bibinfo{pages}{854--859} (\bibinfo{year}{2015}).

\bibitem{Burgwal2017}
\bibinfo{author}{Burgwal, R.} \emph{et~al.}
\newblock \bibinfo{title}{Using an imperfect photonic network to implement
  random unitaries}.
\newblock \emph{\bibinfo{journal}{Optics Express}}
  \textbf{\bibinfo{volume}{25}}, \bibinfo{pages}{28236--28245}
  (\bibinfo{year}{2017}).

\bibitem{Mower2015}
\bibinfo{author}{Mower, J.}, \bibinfo{author}{Harris, N.~C.},
  \bibinfo{author}{Steinbrecher, G.~R.}, \bibinfo{author}{Lahini, Y.} \&
  \bibinfo{author}{Englund, D.}
\newblock \bibinfo{title}{High-fidelity quantum state evolution in imperfect
  photonic integrated circuits}.
\newblock \emph{\bibinfo{journal}{Physical Review A}}
  \textbf{\bibinfo{volume}{92}}, \bibinfo{pages}{032322}
  (\bibinfo{year}{2015}).

\bibitem{Pai2019}
\bibinfo{author}{Pai, S.}, \bibinfo{author}{Bartlett, B.},
  \bibinfo{author}{Solgaard, O.} \& \bibinfo{author}{Miller, D.~A.}
\newblock \bibinfo{title}{Matrix optimization on universal unitary photonic
  devices}.
\newblock \emph{\bibinfo{journal}{Physical Review Applied}}
  \textbf{\bibinfo{volume}{11}}, \bibinfo{pages}{064044}
  (\bibinfo{year}{2019}).

\bibitem{Pai2020}
\bibinfo{author}{Pai, S.} \emph{et~al.}
\newblock \bibinfo{title}{Parallel programming of an arbitrary feedforward
  photonic network}.
\newblock \emph{\bibinfo{journal}{IEEE Journal of Selected Topics in Quantum
  Electronics}}  (\bibinfo{year}{2020}).

\bibitem{Hughes2018}
\bibinfo{author}{Hughes, T.~W.}, \bibinfo{author}{Minkov, M.},
  \bibinfo{author}{Shi, Y.} \& \bibinfo{author}{Fan, S.}
\newblock \bibinfo{title}{Training of photonic neural networks through in situ
  backpropagation and gradient measurement}.
\newblock \emph{\bibinfo{journal}{Optica}} \textbf{\bibinfo{volume}{5}},
  \bibinfo{pages}{864--871} (\bibinfo{year}{2018}).

\bibitem{Miller2013a}
\bibinfo{author}{Miller, D.~A.}
\newblock \bibinfo{title}{Self-aligning universal beam coupler}.
\newblock \emph{\bibinfo{journal}{Optics Express}}
  \textbf{\bibinfo{volume}{21}}, \bibinfo{pages}{6360--6370}
  (\bibinfo{year}{2013}).

\bibitem{Miller2013b}
\bibinfo{author}{Miller, D.~A.}
\newblock \bibinfo{title}{Self-configuring universal linear optical component}.
\newblock \emph{\bibinfo{journal}{Photonics Research}}
  \textbf{\bibinfo{volume}{1}}, \bibinfo{pages}{1--15} (\bibinfo{year}{2013}).

\bibitem{Miller2017}
\bibinfo{author}{Miller, D.~A.}
\newblock \bibinfo{title}{Setting up meshes of interferometers--reversed local
  light interference method}.
\newblock \emph{\bibinfo{journal}{Optics Express}}
  \textbf{\bibinfo{volume}{25}}, \bibinfo{pages}{29233--29248}
  (\bibinfo{year}{2017}).

\bibitem{RyanPaper1}
\bibinfo{author}{Hamerly, R.}, \bibinfo{author}{Bandyopadhyay, S.} \&
  \bibinfo{author}{Englund, D.}
\newblock \bibinfo{title}{Stability of self-configuring large multiport
  interferometers}.
\newblock \emph{\bibinfo{journal}{Physical Review Applied}}
  \textbf{\bibinfo{volume}{18}}, \bibinfo{pages}{024018}
  (\bibinfo{year}{2022}).

\bibitem{RyanPaper2}
\bibinfo{author}{Hamerly, R.}, \bibinfo{author}{Bandyopadhyay, S.} \&
  \bibinfo{author}{Englund, D.}
\newblock \bibinfo{title}{Accurate self-configuration of rectangular multiport
  interferometers}.
\newblock \emph{\bibinfo{journal}{Physical Review Applied}}
  \textbf{\bibinfo{volume}{18}}, \bibinfo{pages}{024019}
  (\bibinfo{year}{2022}).

\bibitem{Annoni2017}
\bibinfo{author}{Annoni, A.} \emph{et~al.}
\newblock \bibinfo{title}{Unscrambling light—automatically undoing strong
  mixing between modes}.
\newblock \emph{\bibinfo{journal}{Light: Science \& Applications}}
  \textbf{\bibinfo{volume}{6}}, \bibinfo{pages}{e17110--e17110}
  (\bibinfo{year}{2017}).

\bibitem{Bandyopadhyay2021}
\bibinfo{author}{Bandyopadhyay, S.}, \bibinfo{author}{Hamerly, R.} \&
  \bibinfo{author}{Englund, D.}
\newblock \bibinfo{title}{Hardware error correction for programmable
  photonics}.
\newblock \emph{\bibinfo{journal}{Optica}} \textbf{\bibinfo{volume}{8}},
  \bibinfo{pages}{1247--1255} (\bibinfo{year}{2021}).

\bibitem{Kumar2021}
\bibinfo{author}{Kumar, S.~P.} \emph{et~al.}
\newblock \bibinfo{title}{Mitigating linear optics imperfections via port
  allocation and compilation}.
\newblock \emph{\bibinfo{journal}{arXiv preprint arXiv:2103.03183}}
  (\bibinfo{year}{2021}).

\bibitem{Lopez2019}
\bibinfo{author}{L{\'o}pez-Pastor, V.~J.}, \bibinfo{author}{Lundeen, J.~S.} \&
  \bibinfo{author}{Marquardt, F.}
\newblock \bibinfo{title}{Arbitrary optical wave evolution with fourier
  transforms and phase masks}.
\newblock \emph{\bibinfo{journal}{Optics Express}}
  \textbf{\bibinfo{volume}{29}}, \bibinfo{pages}{38441--38450}
  (\bibinfo{year}{2021}).

\bibitem{JasvithPaper}
\bibinfo{author}{Basani, J.~R.}, \bibinfo{author}{Vadlamani, S.~K.},
  \bibinfo{author}{Bandyopadhyay, S.}, \bibinfo{author}{Englund, D.~R.} \&
  \bibinfo{author}{Hamerly, R.}
\newblock \bibinfo{title}{A self-similar sine-cosine fractal architecture for
  multiport interferometers}.
\newblock \emph{\bibinfo{journal}{arXiv preprint arXiv:2209.03335}}
  (\bibinfo{year}{2022}).

\bibitem{Miller2015}
\bibinfo{author}{Miller, D.~A.}
\newblock \bibinfo{title}{Perfect optics with imperfect components}.
\newblock \emph{\bibinfo{journal}{Optica}} \textbf{\bibinfo{volume}{2}},
  \bibinfo{pages}{747--750} (\bibinfo{year}{2015}).

\bibitem{Suzuki2015}
\bibinfo{author}{Suzuki, K.} \emph{et~al.}
\newblock \bibinfo{title}{Ultra-high-extinction-ratio 2$\times$2 silicon
  optical switch with variable splitter}.
\newblock \emph{\bibinfo{journal}{Optics Express}}
  \textbf{\bibinfo{volume}{23}}, \bibinfo{pages}{9086--9092}
  (\bibinfo{year}{2015}).

\bibitem{Wilkes2016}
\bibinfo{author}{Wilkes, C.~M.} \emph{et~al.}
\newblock \bibinfo{title}{60 {dB} high-extinction auto-configured
  {M}ach-{Z}ehnder interferometer}.
\newblock \emph{\bibinfo{journal}{Optics Letters}}
  \textbf{\bibinfo{volume}{41}}, \bibinfo{pages}{5318--5321}
  (\bibinfo{year}{2016}).

\bibitem{Wu2019}
\bibinfo{author}{Wu, R.} \emph{et~al.}
\newblock \bibinfo{title}{Fabrication of a multifunctional photonic integrated
  chip on lithium niobate on insulator using femtosecond laser-assisted
  chemomechanical polish}.
\newblock \emph{\bibinfo{journal}{Optics Letters}}
  \textbf{\bibinfo{volume}{44}}, \bibinfo{pages}{4698--4701}
  (\bibinfo{year}{2019}).

\bibitem{Dong2022}
\bibinfo{author}{Dong, M.} \emph{et~al.}
\newblock \bibinfo{title}{High-speed programmable photonic circuits in a
  cryogenically compatible, visible--near-infrared 200 mm {CMOS} architecture}.
\newblock \emph{\bibinfo{journal}{Nature Photonics}}
  \textbf{\bibinfo{volume}{16}}, \bibinfo{pages}{59--65}
  (\bibinfo{year}{2022}).

\bibitem{Bandyopadhyay2022}
\bibinfo{author}{Bandyopadhyay, S.} \emph{et~al.}
\newblock \bibinfo{title}{Single chip photonic deep neural network with
  accelerated training}.
\newblock \emph{\bibinfo{journal}{arXiv preprint arXiv:2208.01623}}
  (\bibinfo{year}{2022}).

\bibitem{Russell2017}
\bibinfo{author}{Russell, N.~J.}, \bibinfo{author}{Chakhmakhchyan, L.},
  \bibinfo{author}{O’Brien, J.~L.} \& \bibinfo{author}{Laing, A.}
\newblock \bibinfo{title}{Direct dialling of {Haar} random unitary matrices}.
\newblock \emph{\bibinfo{journal}{New Journal of Physics}}
  \textbf{\bibinfo{volume}{19}}, \bibinfo{pages}{033007}
  (\bibinfo{year}{2017}).

\bibitem{Haar1933}
\bibinfo{author}{Haar, A.}
\newblock \bibinfo{title}{Der massbegriff in der theorie der kontinuierlichen
  gruppen}.
\newblock \emph{\bibinfo{journal}{Annals of Mathematics}}
  \bibinfo{pages}{147--169} (\bibinfo{year}{1933}).

\bibitem{Tung1985}
\bibinfo{author}{Tung, W.-K.}
\newblock \emph{\bibinfo{title}{Group theory in physics: an introduction to
  symmetry principles, group representations, and special functions in
  classical and quantum physics}} (\bibinfo{publisher}{World Scientific
  Publishing Company}, \bibinfo{year}{1985}).

\bibitem{Wang2020}
\bibinfo{author}{Wang, M.}, \bibinfo{author}{Ribero, A.},
  \bibinfo{author}{Xing, Y.} \& \bibinfo{author}{Bogaerts, W.}
\newblock \bibinfo{title}{Tolerant, broadband tunable 2$\times$2 coupler
  circuit}.
\newblock \emph{\bibinfo{journal}{Optics Express}}
  \textbf{\bibinfo{volume}{28}}, \bibinfo{pages}{5555--5566}
  (\bibinfo{year}{2020}).

\bibitem{Williamson2019}
\bibinfo{author}{Williamson, I.~A.} \emph{et~al.}
\newblock \bibinfo{title}{Reprogrammable electro-optic nonlinear activation
  functions for optical neural networks}.
\newblock \emph{\bibinfo{journal}{IEEE Journal of Selected Topics in Quantum
  Electronics}} \textbf{\bibinfo{volume}{26}}, \bibinfo{pages}{1--12}
  (\bibinfo{year}{2019}).

\bibitem{Fang2019}
\bibinfo{author}{Fang, M. Y.-S.}, \bibinfo{author}{Manipatruni, S.},
  \bibinfo{author}{Wierzynski, C.}, \bibinfo{author}{Khosrowshahi, A.} \&
  \bibinfo{author}{DeWeese, M.~R.}
\newblock \bibinfo{title}{Design of optical neural networks with component
  imprecisions}.
\newblock \emph{\bibinfo{journal}{Optics Express}}
  \textbf{\bibinfo{volume}{27}}, \bibinfo{pages}{14009--14029}
  (\bibinfo{year}{2019}).

\bibitem{Tait2017}
\bibinfo{author}{Tait, A.~N.} \emph{et~al.}
\newblock \bibinfo{title}{Neuromorphic photonic networks using silicon photonic
  weight banks}.
\newblock \emph{\bibinfo{journal}{Scientific Reports}}
  \textbf{\bibinfo{volume}{7}}, \bibinfo{pages}{7430} (\bibinfo{year}{2017}).

\bibitem{Hamerly2019}
\bibinfo{author}{Hamerly, R.}, \bibinfo{author}{Bernstein, L.},
  \bibinfo{author}{Sludds, A.}, \bibinfo{author}{Solja{\v{c}}i{\'c}, M.} \&
  \bibinfo{author}{Englund, D.}
\newblock \bibinfo{title}{Large-scale optical neural networks based on
  photoelectric multiplication}.
\newblock \emph{\bibinfo{journal}{Physical Review X}}
  \textbf{\bibinfo{volume}{9}}, \bibinfo{pages}{021032} (\bibinfo{year}{2019}).

\bibitem{Bernstein2022}
\bibinfo{author}{Bernstein, L.} \emph{et~al.}
\newblock \bibinfo{title}{Single-shot optical neural network}.
\newblock \emph{\bibinfo{journal}{arXiv preprint arXiv:2205.09103}}
  (\bibinfo{year}{2022}).

\bibitem{Chen2022}
\bibinfo{author}{Chen, Z.} \emph{et~al.}
\newblock \bibinfo{title}{Deep learning with coherent VCSEL neural networks}.
\newblock \emph{\bibinfo{journal}{arXiv preprint arXiv:2207.05329}}
  (\bibinfo{year}{2022}).

\bibitem{LeCun1998}
\bibinfo{author}{LeCun, Y.}, \bibinfo{author}{Bottou, L.},
  \bibinfo{author}{Bengio, Y.} \& \bibinfo{author}{Haffner, P.}
\newblock \bibinfo{title}{Gradient-based learning applied to document
  recognition}.
\newblock \emph{\bibinfo{journal}{Proceedings of the IEEE}}
  \textbf{\bibinfo{volume}{86}}, \bibinfo{pages}{2278--2324}
  (\bibinfo{year}{1998}).

\bibitem{Neurophox}
\bibinfo{author}{Pai, S.}
\newblock \bibinfo{title}{Neurophox: a simulation framework for unitary neural
  networks and photonic devices}.
\newblock \bibinfo{howpublished}{Online at:
  \url{https://github.com/solgaardlab/neurophox}} (\bibinfo{year}{2020}).

\bibitem{Mikkelsen2014}
\bibinfo{author}{Mikkelsen, J.~C.}, \bibinfo{author}{Sacher, W.~D.} \&
  \bibinfo{author}{Poon, J.~K.}
\newblock \bibinfo{title}{Dimensional variation tolerant silicon-on-insulator
  directional couplers}.
\newblock \emph{\bibinfo{journal}{Optics Express}}
  \textbf{\bibinfo{volume}{22}}, \bibinfo{pages}{3145--3150}
  (\bibinfo{year}{2014}).

\bibitem{Soldano1995}
\bibinfo{author}{Soldano, L.~B.} \& \bibinfo{author}{Pennings, E.~C.}
\newblock \bibinfo{title}{Optical multi-mode interference devices based on
  self-imaging: principles and applications}.
\newblock \emph{\bibinfo{journal}{Journal of Lightwave Technology}}
  \textbf{\bibinfo{volume}{13}}, \bibinfo{pages}{615--627}
  (\bibinfo{year}{1995}).

\bibitem{Maese2013}
\bibinfo{author}{Maese-Novo, A.} \emph{et~al.}
\newblock \bibinfo{title}{Wavelength independent multimode interference
  coupler}.
\newblock \emph{\bibinfo{journal}{Optics Express}}
  \textbf{\bibinfo{volume}{21}}, \bibinfo{pages}{7033--7040}
  (\bibinfo{year}{2013}).

\bibitem{Wang2016}
\bibinfo{author}{Wang, Y.} \emph{et~al.}
\newblock \bibinfo{title}{Compact broadband directional couplers using
  subwavelength gratings}.
\newblock \emph{\bibinfo{journal}{IEEE Photonics Journal}}
  \textbf{\bibinfo{volume}{8}}, \bibinfo{pages}{1--8} (\bibinfo{year}{2016}).

\bibitem{Ye2020}
\bibinfo{author}{Ye, C.} \& \bibinfo{author}{Dai, D.}
\newblock \bibinfo{title}{Ultra-compact broadband 2$\times$2 3 {dB} power
  splitter using a subwavelength-grating-assisted asymmetric directional
  coupler}.
\newblock \emph{\bibinfo{journal}{Journal of Lightwave Technology}}
  \textbf{\bibinfo{volume}{38}}, \bibinfo{pages}{2370--2375}
  (\bibinfo{year}{2020}).

\bibitem{Morino2014}
\bibinfo{author}{Morino, H.}, \bibinfo{author}{Maruyama, T.} \&
  \bibinfo{author}{Iiyama, K.}
\newblock \bibinfo{title}{Reduction of wavelength dependence of coupling
  characteristics using {Si} optical waveguide curved directional coupler}.
\newblock \emph{\bibinfo{journal}{Journal of Lightwave Technology}}
  \textbf{\bibinfo{volume}{32}}, \bibinfo{pages}{2188--2192}
  (\bibinfo{year}{2014}).

\bibitem{Lu2015}
\bibinfo{author}{Lu, Z.} \emph{et~al.}
\newblock \bibinfo{title}{Broadband silicon photonic directional coupler using
  asymmetric-waveguide based phase control}.
\newblock \emph{\bibinfo{journal}{Optics Express}}
  \textbf{\bibinfo{volume}{23}}, \bibinfo{pages}{3795--3808}
  (\bibinfo{year}{2015}).

\bibitem{Bogaerts2019}
\bibinfo{author}{Bogaerts, W.}, \bibinfo{author}{Xing, Y.} \&
  \bibinfo{author}{Khan, U.}
\newblock \bibinfo{title}{Layout-aware variability analysis, yield prediction,
  and optimization in photonic integrated circuits}.
\newblock \emph{\bibinfo{journal}{IEEE Journal of Selected Topics in Quantum
  Electronics}} \textbf{\bibinfo{volume}{25}}, \bibinfo{pages}{1--13}
  (\bibinfo{year}{2019}).

\bibitem{Suzuki2018}
\bibinfo{author}{Suzuki, K.} \emph{et~al.}
\newblock \bibinfo{title}{Low-insertion-loss and power-efficient 32$\times$32
  silicon photonics switch with extremely high-$\delta$ silica {PLC}
  connector}.
\newblock \emph{\bibinfo{journal}{Journal of Lightwave Technology}}
  \textbf{\bibinfo{volume}{37}}, \bibinfo{pages}{116--122}
  (\bibinfo{year}{2018}).

\bibitem{Feldmann2021}
\bibinfo{author}{Feldmann, J.} \emph{et~al.}
\newblock \bibinfo{title}{Parallel convolutional processing using an integrated
  photonic tensor core}.
\newblock \emph{\bibinfo{journal}{Nature}} \textbf{\bibinfo{volume}{589}},
  \bibinfo{pages}{52--58} (\bibinfo{year}{2021}).

\bibitem{Xu2021}
\bibinfo{author}{Xu, X.} \emph{et~al.}
\newblock \bibinfo{title}{11 TOPS photonic convolutional accelerator for
  optical neural networks}.
\newblock \emph{\bibinfo{journal}{Nature}} \textbf{\bibinfo{volume}{589}},
  \bibinfo{pages}{44--51} (\bibinfo{year}{2021}).

\bibitem{Sludds2022}
\bibinfo{author}{Sludds, A.} \emph{et~al.}
\newblock \bibinfo{title}{Delocalized photonic deep learning on the internet's
  edge}.
\newblock \emph{\bibinfo{journal}{Science}}
  \textbf{\bibinfo{volume}{378}}, \bibinfo{pages}{270--276} (\bibinfo{year}{2022}).

\bibitem{Davis2022}
\bibinfo{author}{Davis~III, R.}, \bibinfo{author}{Chen, Z.},
  \bibinfo{author}{Hamerly, R.} \& \bibinfo{author}{Englund, D.}
\newblock \bibinfo{title}{Frequency-encoded deep learning with speed-of-light
  dominated latency}.
\newblock \emph{\bibinfo{journal}{arXiv preprint arXiv:2207.06883}}
  (\bibinfo{year}{2022}).

\bibitem{Fukazawa2004}
\bibinfo{author}{Fukazawa, T.}, \bibinfo{author}{Hirano, T.},
  \bibinfo{author}{Ohno, F.} \& \bibinfo{author}{Baba, T.}
\newblock \bibinfo{title}{Low loss intersection of {Si} photonic wire
  waveguides}.
\newblock \emph{\bibinfo{journal}{Japanese Journal of Applied Physics}}
  \textbf{\bibinfo{volume}{43}}, \bibinfo{pages}{646} (\bibinfo{year}{2004}).

\bibitem{Chen2006}
\bibinfo{author}{Chen, H.} \& \bibinfo{author}{Poon, A.~W.}
\newblock \bibinfo{title}{Low-loss multimode-interference-based crossings for
  silicon wire waveguides}.
\newblock \emph{\bibinfo{journal}{IEEE Photonics Technology Letters}}
  \textbf{\bibinfo{volume}{18}}, \bibinfo{pages}{2260--2262}
  (\bibinfo{year}{2006}).

\bibitem{Ma2013}
\bibinfo{author}{Ma, Y.} \emph{et~al.}
\newblock \bibinfo{title}{Ultralow loss single layer submicron silicon
  waveguide crossing for {SOI} optical interconnect}.
\newblock \emph{\bibinfo{journal}{Optics Express}}
  \textbf{\bibinfo{volume}{21}}, \bibinfo{pages}{29374--29382}
  (\bibinfo{year}{2013}).

\bibitem{Dumais2017}
\bibinfo{author}{Dumais, P.}, \bibinfo{author}{Goodwill, D.},
  \bibinfo{author}{Celo, D.}, \bibinfo{author}{Jiang, J.} \&
  \bibinfo{author}{Bernier, E.}
\newblock \bibinfo{title}{Three-mode synthesis of slab gaussian beam in
  ultra-low-loss in-plane nanophotonic silicon waveguide crossing}.
\newblock In \emph{\bibinfo{booktitle}{2017 IEEE 14th International Conference
  on Group IV Photonics (GFP)}}, \bibinfo{pages}{97--98}
  (\bibinfo{organization}{IEEE}, \bibinfo{year}{2017}).

\bibitem{Wu2020}
\bibinfo{author}{Wu, S.}, \bibinfo{author}{Mu, X.}, \bibinfo{author}{Cheng,
  L.}, \bibinfo{author}{Mao, S.} \& \bibinfo{author}{Fu, H.}
\newblock \bibinfo{title}{State-of-the-art and perspectives on silicon
  waveguide crossings: a review}.
\newblock \emph{\bibinfo{journal}{Micromachines}}
  \textbf{\bibinfo{volume}{11}}, \bibinfo{pages}{326} (\bibinfo{year}{2020}).

\newpage

\bibitem{SriPaper}
\bibinfo{author}{Vadlamani, S.~K.}, \bibinfo{author}{Englund, D.} \&
  \bibinfo{author}{Hamerly, R.}
\newblock \bibinfo{title}{Transferable learning on analog hardware}.
\newblock \emph{\bibinfo{journal}{arXiv preprint arXiv:2210.06632}}
  (\bibinfo{year}{2022}).

\bibitem{Brown2004}
\bibinfo{author}{Brown, K.~R.}, \bibinfo{author}{Harrow, A.~W.} \&
  \bibinfo{author}{Chuang, I.~L.}
\newblock \bibinfo{title}{Arbitrarily accurate composite pulse sequences}.
\newblock \emph{\bibinfo{journal}{Physical Review A}}
  \textbf{\bibinfo{volume}{70}}, \bibinfo{pages}{052318}
  (\bibinfo{year}{2004}).

\bibitem{Bulmer2020}
\bibinfo{author}{Bulmer, J.}, \bibinfo{author}{Jones, J.} \&
  \bibinfo{author}{Walmsley, I.}
\newblock \bibinfo{title}{Drive-noise tolerant optical switching inspired by
  composite pulses}.
\newblock \emph{\bibinfo{journal}{Optics Express}}
  \textbf{\bibinfo{volume}{28}}, \bibinfo{pages}{8646--8657}
  (\bibinfo{year}{2020}).

\bibitem{Little1997}
\bibinfo{author}{Little, B.~E.} \& \bibinfo{author}{Murphy, T.}
\newblock \bibinfo{title}{Design rules for maximally flat
  wavelength-insensitive optical power dividers using {M}ach-{Z}ehnder
  structures}.
\newblock \emph{\bibinfo{journal}{IEEE Photonics Technology Letters}}
  \textbf{\bibinfo{volume}{9}}, \bibinfo{pages}{1607--1609}
  (\bibinfo{year}{1997}).

\bibitem{Bell2021}
\bibinfo{author}{Bell, B.~A.} \& \bibinfo{author}{Walmsley, I.~A.}
\newblock \bibinfo{title}{Further compactifying linear optical unitaries}.
\newblock \emph{\bibinfo{journal}{APL Photonics}} \textbf{\bibinfo{volume}{6}},
  \bibinfo{pages}{070804} (\bibinfo{year}{2021}).

\bibitem{Meshes}
\bibinfo{author}{Hamerly, R.}
\newblock \bibinfo{title}{Meshes: tools for modeling photonic beamsplitter mesh
  networks}.
\newblock \bibinfo{howpublished}{Online at:
  \url{https://github.com/QPG-MIT/meshes}} (\bibinfo{year}{2021}).

\end{thebibliography}

\begin{thebibliography}{10}
\expandafter\ifx\csname url\endcsname\relax
  \def\url#1{\texttt{#1}}\fi
\expandafter\ifx\csname urlprefix\endcsname\relax\def\urlprefix{URL }\fi
\providecommand{\bibinfo}[2]{#2}
\providecommand{\eprint}[2][]{\url{#2}}

\bibitem{S_Reck1994}
\bibinfo{author}{Reck, M.}, \bibinfo{author}{Zeilinger, A.},
  \bibinfo{author}{Bernstein, H.~J.} \& \bibinfo{author}{Bertani, P.}
\newblock \bibinfo{title}{Experimental realization of any discrete unitary
  operator}.
\newblock \emph{\bibinfo{journal}{Physical Review Letters}}
  \textbf{\bibinfo{volume}{73}}, \bibinfo{pages}{58} (\bibinfo{year}{1994}).

\bibitem{S_Clements2016}
\bibinfo{author}{Clements, W.~R.}, \bibinfo{author}{Humphreys, P.~C.},
  \bibinfo{author}{Metcalf, B.~J.}, \bibinfo{author}{Kolthammer, W.~S.} \&
  \bibinfo{author}{Walmsley, I.~A.}
\newblock \bibinfo{title}{Optimal design for universal multiport
  interferometers}.
\newblock \emph{\bibinfo{journal}{Optica}} \textbf{\bibinfo{volume}{3}},
  \bibinfo{pages}{1460--1465} (\bibinfo{year}{2016}).

\bibitem{S_Miller2013b}
\bibinfo{author}{Miller, D.~A.}
\newblock \bibinfo{title}{Self-configuring universal linear optical component}.
\newblock \emph{\bibinfo{journal}{Photonics Research}}
  \textbf{\bibinfo{volume}{1}}, \bibinfo{pages}{1--15} (\bibinfo{year}{2013}).

\bibitem{S_Annoni2017}
\bibinfo{author}{Annoni, A.} \emph{et~al.}
\newblock \bibinfo{title}{Unscrambling light—automatically undoing strong
  mixing between modes}.
\newblock \emph{\bibinfo{journal}{Light: Science \& Applications}}
  \textbf{\bibinfo{volume}{6}}, \bibinfo{pages}{e17110--e17110}
  (\bibinfo{year}{2017}).

\bibitem{S_Miller2017}
\bibinfo{author}{Miller, D.~A.}
\newblock \bibinfo{title}{Setting up meshes of interferometers--reversed local
  light interference method}.
\newblock \emph{\bibinfo{journal}{Optics Express}}
  \textbf{\bibinfo{volume}{25}}, \bibinfo{pages}{29233--29248}
  (\bibinfo{year}{2017}).

\bibitem{S_Hamerly2022}
\bibinfo{author}{Hamerly, R.}, \bibinfo{author}{Bandyopadhyay, S.} \&
  \bibinfo{author}{Englund, D.}
\newblock \bibinfo{title}{Accurate self-configuration of rectangular multiport
  interferometers}.
\newblock \emph{\bibinfo{journal}{Physical Review Applied}}
  \textbf{\bibinfo{volume}{18}}, \bibinfo{pages}{024019}
  (\bibinfo{year}{2022}).

\bibitem{S_Hamerly2022a}
\bibinfo{author}{Hamerly, R.}, \bibinfo{author}{Bandyopadhyay, S.} \&
  \bibinfo{author}{Englund, D.}
\newblock \bibinfo{title}{Stability of self-configuring large multiport
  interferometers}.
\newblock \emph{\bibinfo{journal}{Physical Review Applied}}
  \textbf{\bibinfo{volume}{18}}, \bibinfo{pages}{024018}
  (\bibinfo{year}{2022}).

\bibitem{S_Bandyopadhyay2022}
\bibinfo{author}{Bandyopadhyay, S.} \emph{et~al.}
\newblock \bibinfo{title}{Single chip photonic deep neural network with
  accelerated training}.
\newblock \emph{\bibinfo{journal}{arXiv preprint arXiv:2208.01623}}
  (\bibinfo{year}{2022}).

\bibitem{S_Mower2015}
\bibinfo{author}{Mower, J.}, \bibinfo{author}{Harris, N.~C.},
  \bibinfo{author}{Steinbrecher, G.~R.}, \bibinfo{author}{Lahini, Y.} \&
  \bibinfo{author}{Englund, D.}
\newblock \bibinfo{title}{High-fidelity quantum state evolution in imperfect
  photonic integrated circuits}.
\newblock \emph{\bibinfo{journal}{Physical Review A}}
  \textbf{\bibinfo{volume}{92}}, \bibinfo{pages}{032322}
  (\bibinfo{year}{2015}).

\bibitem{S_Burgwal2017}
\bibinfo{author}{Burgwal, R.} \emph{et~al.}
\newblock \bibinfo{title}{Using an imperfect photonic network to implement
  random unitaries}.
\newblock \emph{\bibinfo{journal}{Optics Express}}
  \textbf{\bibinfo{volume}{25}}, \bibinfo{pages}{28236--28245}
  (\bibinfo{year}{2017}).

\bibitem{S_Pai2019}
\bibinfo{author}{Pai, S.}, \bibinfo{author}{Bartlett, B.},
  \bibinfo{author}{Solgaard, O.} \& \bibinfo{author}{Miller, D.~A.}
\newblock \bibinfo{title}{Matrix optimization on universal unitary photonic
  devices}.
\newblock \emph{\bibinfo{journal}{Physical Review Applied}}
  \textbf{\bibinfo{volume}{11}}, \bibinfo{pages}{064044}
  (\bibinfo{year}{2019}).

\bibitem{S_Hughes2018}
\bibinfo{author}{Hughes, T.~W.}, \bibinfo{author}{Minkov, M.},
  \bibinfo{author}{Shi, Y.} \& \bibinfo{author}{Fan, S.}
\newblock \bibinfo{title}{Training of photonic neural networks through in situ
  backpropagation and gradient measurement}.
\newblock \emph{\bibinfo{journal}{Optica}} \textbf{\bibinfo{volume}{5}},
  \bibinfo{pages}{864--871} (\bibinfo{year}{2018}).

\bibitem{S_Neurophox}
\bibinfo{author}{Pai, S.}
\newblock \bibinfo{title}{Neurophox: a simulation framework for unitary neural
  networks and photonic devices}.
\newblock \bibinfo{howpublished}{Online at:
  \url{https://github.com/solgaardlab/neurophox}} (\bibinfo{year}{2020}).

\bibitem{S_Meshes}
\bibinfo{author}{Hamerly, R.}
\newblock \bibinfo{title}{Meshes: tools for modeling photonic beamsplitter mesh
  networks}.
\newblock \bibinfo{howpublished}{Online at:
  \url{https://github.com/QPG-MIT/meshes}} (\bibinfo{year}{2021}).

\bibitem{S_Yang2015}
\bibinfo{author}{Yang, Y.} \emph{et~al.}
\newblock \bibinfo{title}{Phase coherence length in silicon photonic platform}.
\newblock \emph{\bibinfo{journal}{Optics Express}}
  \textbf{\bibinfo{volume}{23}}, \bibinfo{pages}{16890--16902}
  (\bibinfo{year}{2015}).

\bibitem{S_Chrostowski2014}
\bibinfo{author}{Chrostowski, L.} \emph{et~al.}
\newblock \bibinfo{title}{Impact of fabrication non-uniformity on chip-scale
  silicon photonic integrated circuits}.
\newblock In \emph{\bibinfo{booktitle}{Optical Fiber Communication
  Conference}}, \bibinfo{pages}{Th2A--37} (\bibinfo{organization}{Optical
  Society of America}, \bibinfo{year}{2014}).

\bibitem{S_Bogaerts2019}
\bibinfo{author}{Bogaerts, W.}, \bibinfo{author}{Xing, Y.} \&
  \bibinfo{author}{Khan, U.}
\newblock \bibinfo{title}{Layout-aware variability analysis, yield prediction,
  and optimization in photonic integrated circuits}.
\newblock \emph{\bibinfo{journal}{IEEE Journal of Selected Topics in Quantum
  Electronics}} \textbf{\bibinfo{volume}{25}}, \bibinfo{pages}{1--13}
  (\bibinfo{year}{2019}).

\bibitem{S_Kawachi1993}
\bibinfo{author}{Kawachi, M.} \emph{et~al.}
\newblock \bibinfo{title}{Silica-based optical-matrix switch with intersecting
  {M}ach-{Z}ehnder waveguides for larger fabrication tolerances}.
\newblock In \emph{\bibinfo{booktitle}{Optical Fiber Communication
  Conference}}, \bibinfo{pages}{TuH4} (\bibinfo{organization}{Optical Society
  of America}, \bibinfo{year}{1993}).

\bibitem{S_Nagase1994}
\bibinfo{author}{Nagase, R.} \emph{et~al.}
\newblock \bibinfo{title}{Silica-based 8$\times$8 optical matrix switch module
  with hybrid integrated driving circuits and its system application}.
\newblock \emph{\bibinfo{journal}{Journal of Lightwave Technology}}
  \textbf{\bibinfo{volume}{12}}, \bibinfo{pages}{1631--1639}
  (\bibinfo{year}{1994}).

\bibitem{S_Goh1999}
\bibinfo{author}{Goh, T.}, \bibinfo{author}{Himeno, A.},
  \bibinfo{author}{Okuno, M.}, \bibinfo{author}{Takahashi, H.} \&
  \bibinfo{author}{Hattori, K.}
\newblock \bibinfo{title}{High-extinction ratio and low-loss silica-based 88
  strictly nonblocking thermooptic matrix switch}.
\newblock \emph{\bibinfo{journal}{Journal of Lightwave Technology}}
  \textbf{\bibinfo{volume}{17}}, \bibinfo{pages}{1192} (\bibinfo{year}{1999}).

\bibitem{S_Okuno1999}
\bibinfo{author}{Okuno, M.} \emph{et~al.}
\newblock \bibinfo{title}{Silica-based 8$\times$8 optical matrix switch
  integrating new switching units with large fabrication tolerance}.
\newblock \emph{\bibinfo{journal}{Journal of Lightwave Technology}}
  \textbf{\bibinfo{volume}{17}}, \bibinfo{pages}{771--781}
  (\bibinfo{year}{1999}).

\bibitem{S_Shoji2010}
\bibinfo{author}{Shoji, Y.} \emph{et~al.}
\newblock \bibinfo{title}{Low-crosstalk 2$\times$2 thermo-optic switch with
  silicon wire waveguides}.
\newblock \emph{\bibinfo{journal}{Optics Express}}
  \textbf{\bibinfo{volume}{18}}, \bibinfo{pages}{9071--9075}
  (\bibinfo{year}{2010}).

\bibitem{S_Harris2017}
\bibinfo{author}{Harris, N.~C.} \emph{et~al.}
\newblock \bibinfo{title}{Quantum transport simulations in a programmable
  nanophotonic processor}.
\newblock \emph{\bibinfo{journal}{Nature Photonics}}
  \textbf{\bibinfo{volume}{11}}, \bibinfo{pages}{447--452}
  (\bibinfo{year}{2017}).

\bibitem{S_Dumais2017b}
\bibinfo{author}{Dumais, P.} \emph{et~al.}
\newblock \bibinfo{title}{Silicon photonic switch subsystem with 900
  monolithically integrated calibration photodiodes and 64-fiber package}.
\newblock \emph{\bibinfo{journal}{Journal of Lightwave Technology}}
  \textbf{\bibinfo{volume}{36}}, \bibinfo{pages}{233--238}
  (\bibinfo{year}{2017}).

\bibitem{S_Suzuki2018}
\bibinfo{author}{Suzuki, K.} \emph{et~al.}
\newblock \bibinfo{title}{Low-insertion-loss and power-efficient 32$\times$32
  silicon photonics switch with extremely high-$\delta$ silica {PLC}
  connector}.
\newblock \emph{\bibinfo{journal}{Journal of Lightwave Technology}}
  \textbf{\bibinfo{volume}{37}}, \bibinfo{pages}{116--122}
  (\bibinfo{year}{2018}).

\bibitem{S_Wilkes2016}
\bibinfo{author}{Wilkes, C.~M.} \emph{et~al.}
\newblock \bibinfo{title}{60 {dB} high-extinction auto-configured
  {M}ach-{Z}ehnder interferometer}.
\newblock \emph{\bibinfo{journal}{Optics Letters}}
  \textbf{\bibinfo{volume}{41}}, \bibinfo{pages}{5318--5321}
  (\bibinfo{year}{2016}).

\bibitem{S_Dong2022}
\bibinfo{author}{Dong, M.} \emph{et~al.}
\newblock \bibinfo{title}{High-speed programmable photonic circuits in a
  cryogenically compatible, visible--near-infrared 200 mm {CMOS} architecture}.
\newblock \emph{\bibinfo{journal}{Nature Photonics}}
  \textbf{\bibinfo{volume}{16}}, \bibinfo{pages}{59--65}
  (\bibinfo{year}{2022}).

\bibitem{S_Suzuki2015}
\bibinfo{author}{Suzuki, K.} \emph{et~al.}
\newblock \bibinfo{title}{Ultra-high-extinction-ratio 2$\times$2 silicon
  optical switch with variable splitter}.
\newblock \emph{\bibinfo{journal}{Optics Express}}
  \textbf{\bibinfo{volume}{23}}, \bibinfo{pages}{9086--9092}
  (\bibinfo{year}{2015}).

\bibitem{S_Orlandi2013}
\bibinfo{author}{Orlandi, P.} \emph{et~al.}
\newblock \bibinfo{title}{Tunable silicon photonics directional coupler driven
  by a transverse temperature gradient}.
\newblock \emph{\bibinfo{journal}{Optics Letters}}
  \textbf{\bibinfo{volume}{38}}, \bibinfo{pages}{863--865}
  (\bibinfo{year}{2013}).

\bibitem{S_Miller2015}
\bibinfo{author}{Miller, D.~A.}
\newblock \bibinfo{title}{Perfect optics with imperfect components}.
\newblock \emph{\bibinfo{journal}{Optica}} \textbf{\bibinfo{volume}{2}},
  \bibinfo{pages}{747--750} (\bibinfo{year}{2015}).

\bibitem{S_Feldmann2021}
\bibinfo{author}{Feldmann, J.} \emph{et~al.}
\newblock \bibinfo{title}{Parallel convolutional processing using an integrated
  photonic tensor core}.
\newblock \emph{\bibinfo{journal}{Nature}} \textbf{\bibinfo{volume}{589}},
  \bibinfo{pages}{52--58} (\bibinfo{year}{2021}).

\bibitem{S_Wang2020}
\bibinfo{author}{Wang, M.}, \bibinfo{author}{Ribero, A.},
  \bibinfo{author}{Xing, Y.} \& \bibinfo{author}{Bogaerts, W.}
\newblock \bibinfo{title}{Tolerant, broadband tunable 2$\times$2 coupler
  circuit}.
\newblock \emph{\bibinfo{journal}{Optics Express}}
  \textbf{\bibinfo{volume}{28}}, \bibinfo{pages}{5555--5566}
  (\bibinfo{year}{2020}).
  
\bibitem{S_Taballione2022}
\bibinfo{author}{Taballione, C.} \emph{et~al.}
\newblock \bibinfo{title}{20-mode universal quantum photonic processor}.
\newblock \emph{\bibinfo{journal}{arXiv preprint arXiv:2203.01801}}
  (\bibinfo{year}{2022}).

\newpage

\bibitem{S_Wu2019}
\bibinfo{author}{Wu, R.} \emph{et~al.}
\newblock \bibinfo{title}{Fabrication of a multifunctional photonic integrated
  chip on lithium niobate on insulator using femtosecond laser-assisted
  chemomechanical polish}.
\newblock \emph{\bibinfo{journal}{Optics Letters}}
  \textbf{\bibinfo{volume}{44}}, \bibinfo{pages}{4698--4701}
  (\bibinfo{year}{2019}).

\bibitem{S_Siew2021}
\bibinfo{author}{Siew, S.~Y.} \emph{et~al.}
\newblock \bibinfo{title}{Review of silicon photonics technology and platform
  development}.
\newblock \emph{\bibinfo{journal}{Journal of Lightwave Technology}}
  \textbf{\bibinfo{volume}{39}}, \bibinfo{pages}{4374--4389}
  (\bibinfo{year}{2021}).

\bibitem{S_Kumar2021}
\bibinfo{author}{Kumar, S.~P.} \emph{et~al.}
\newblock \bibinfo{title}{Mitigating linear optics imperfections via port
  allocation and compilation}.
\newblock \emph{\bibinfo{journal}{arXiv preprint arXiv:2103.03183}}
  (\bibinfo{year}{2021}).

\bibitem{S_AlQadasi2022}
\bibinfo{author}{Al-Qadasi, M.}, \bibinfo{author}{Chrostowski, L.},
  \bibinfo{author}{Shastri, B.} \& \bibinfo{author}{Shekhar, S.}
\newblock \bibinfo{title}{Scaling up silicon photonic-based accelerators:
  Challenges and opportunities}.
\newblock \emph{\bibinfo{journal}{APL Photonics}} \textbf{\bibinfo{volume}{7}},
  \bibinfo{pages}{020902} (\bibinfo{year}{2022}).

\bibitem{S_Shen2017}
\bibinfo{author}{Shen, Y.} \emph{et~al.}
\newblock \bibinfo{title}{Deep learning with coherent nanophotonic circuits}.
\newblock \emph{\bibinfo{journal}{Nature Photonics}}
  \textbf{\bibinfo{volume}{11}}, \bibinfo{pages}{441} (\bibinfo{year}{2017}).

\bibitem{S_Qiu2020}
\bibinfo{author}{Qiu, H.} \emph{et~al.}
\newblock \bibinfo{title}{Energy-efficient thermo-optic silicon phase shifter
  with well-balanced overall performance}.
\newblock \emph{\bibinfo{journal}{Optics Letters}}
  \textbf{\bibinfo{volume}{45}}, \bibinfo{pages}{4806--4809}
  (\bibinfo{year}{2020}).

\end{thebibliography}


\section*{Acknowledgements}

S.B.\ is supported by an NSF Graduate Research Fellowship.  D.E.\ acknowledges funding from AFOSR (no.\ FA9550-20-1-0113, FA9550-16-1-0391).  The authors thank Prof.\ David A.\ B.\ Miller and Dr. Sunil Pai for helpful discussions.

\section*{Author contributions}

S.B.\ and R.H.\ jointly conceived the idea.  R.H.\ developed the theory, performed the simulations and data analysis, and wrote the manuscript.  All authors contributed to discussion of the results.

\section*{Competing interests}

R.H., S.B., and D.E.\ are inventors on patent applications No.\ 63/151,103 and 63/196,301 describing methods for self-configuration and error correction in linear photonic circuits.

\newpage

\setcounter{equation}{0}
\setcounter{figure}{0}
\setcounter{table}{0}
\makeatletter
\renewcommand{\thetable}{S\arabic{table}}
\renewcommand{\thefigure}{S\arabic{figure}}
\renewcommand{\thesection}{S\arabic{section}}
\renewcommand{\theequation}{S\arabic{equation}}
\renewcommand{\thefigure}{S\arabic{figure}}
\renewcommand{\bibnumfmt}[1]{[S#1]}
\renewcommand{\citenumfont}[1]{S#1}

\newcommand{\EqII}{\ref{eq:ii}}
\newcommand{\EqIII}{\ref{eq:ec3mzi2}}
\newcommand{\EqIV}{\ref{eq:iv}}
\newcommand{\EqXI}{\ref{eq:cov0}}  
\newcommand{\EqXII}{\ref{eq:ec}}
\newcommand{\EqXVII}{\ref{eq:ec0}}
\newcommand{\EqXIX}{\ref{eq:res3}}
\newcommand{\EqXX}{\ref{eq:ec3mzi}}
\newcommand{\EqXXIV}{\ref{eq:resX}}
\newcommand{\EqXXV}{\ref{eq:ecmzix}}
\newcommand{\TabII}{\ref{tab:t2}}
\newcommand{\FigVI}{\ref{fig:f8}}


\section*{Supplementary Material}

\section{Error Correction Methods}
\label{sec:si-ecm}

Correction of hardware errors is performed using the nulling method, which is based on the diagonalization of a unitary matrix using Givens rotations.  This is closely related to the QR decomposition for the Reck triangle \cite{S_Reck1994}, and a related decomposition for the more compact Clements rectangle \cite{S_Clements2016}.  The original nulling proposal was restricted to triangular (Reck) meshes and used internal tap detectors to monitor the output power of each MZI \cite{S_Miller2013b, S_Annoni2017}.  Subsequently, the method was extended to generic mesh types \cite{S_Miller2017}, and Ref.~\cite{S_Hamerly2022} showed that that external detectors are sufficient for both Reck and Clements meshes.

\begin{figure}[bp]
\begin{center}
\includegraphics[width=1.00\columnwidth]{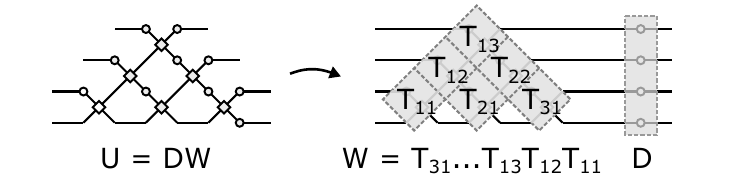}
\caption{Reck decomposition of a $4\times 4$ programmable unitary.}
\label{fig:fs3}
\end{center}
\end{figure}

Following Appendix~A of Ref.~\cite{S_Hamerly2022}, we describe here the nulling procedure for configuring a Reck mesh.  First, we write the coupling matrix for the multiport interferometer as a product of the $2\times 2$ MZI blocks and an external phase screen:
\beq
	U = D \underbrace{\bigl(T_{N-1,1} \ldots T_{13} T_{12} T_{11}\bigr)}_{W} \label{eq:si-udw}
\eeq
Here, the $T_{mn}$ represent tunable couplers (MZI, 3-MZI, MZI+X, etc.) while $D$ is a diagonal matrix encoding the output phase shifts.  The $T_{mn}$ are ordered along rising diagonals as shown in Fig.~\ref{fig:fs3} (nulling also works on falling diagonals \cite{S_Hamerly2022}). 

\begin{figure}[b!]
\begin{center}
\includegraphics[width=1.00\columnwidth]{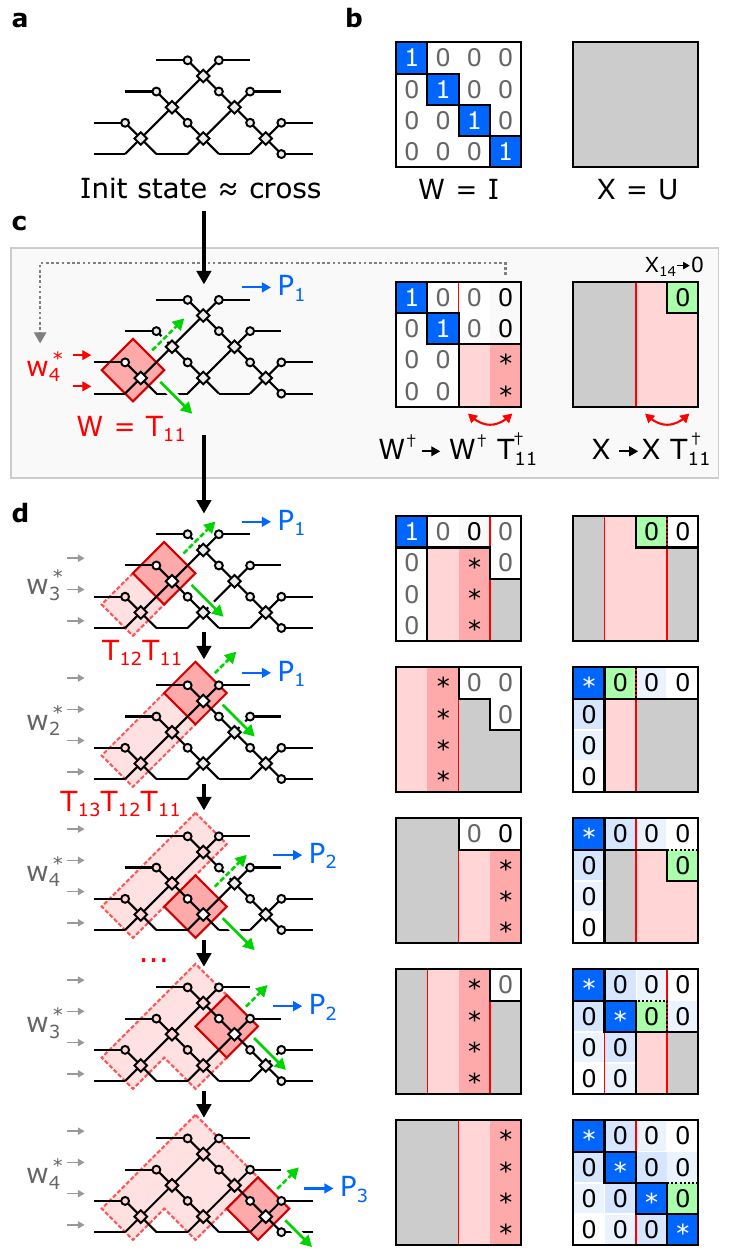}
\caption{Configuration of a $4\times 4$ Reck mesh by measurement-assisted nulling, following the procedure of Ref.~\cite{S_Hamerly2022}.}
\label{fig:fs4}
\end{center}
\end{figure}

Fig.~\ref{fig:fs4} traces out the nulling steps for a $4\times 4$ Reck mesh.  We start by initializing the mesh to approximately the cross state, Fig.~\ref{fig:fs4}(a).  We keep track of two matrices (Fig.~\ref{fig:fs4}(b)): $W = T_{N-1,1} \ldots T_{11}$ is the partial product of all configured MZIs, and $X = U W^\dagger$, where $U$ is the target unitary.  At the beginning, none of the MZIs are configured, so $W = I$ and $X = U$.  At each step, with an example shown in Fig.~\ref{fig:fs4}(c), we configure target MZI $T_{mn}$, which updates $W$ and $X$ by Givens rotations $W \rightarrow T_{mn} W$, $X \rightarrow X T_{mn}^\dagger$.  The target $T_{mn}$ is chosen to zero a particular off-diagonal element $X_{ij}$.  Subsequent MZIs are configured in a sequence that successively zeroes off-diagonal elements of $X$ (Fig.~\ref{fig:fs4}(d)).  If all MZIs are configured properly, at the end of the procedure, $X$ is diagonalized so $U = DW$, and the output phases (elements of $D$) can be read off by inspection.

\begin{figure}[t]
\begin{center}
\includegraphics[width=1.00\columnwidth]{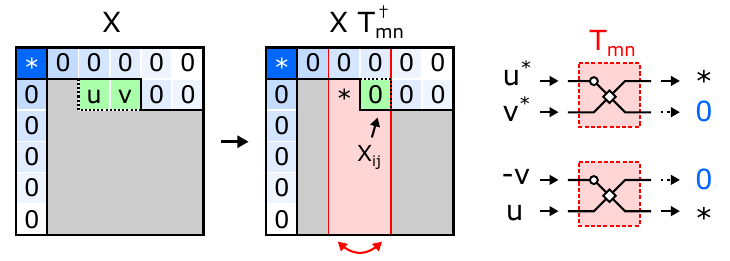}
\caption{Right-multiplication by $T_m^\dagger$ mixes the elements $(u, v)$ of $X$ and zeroes out the rightmost one.  This is equivalent to the condition Eq.~(\ref{eq:si-tuv}).}
\label{fig:fs5}
\end{center}
\end{figure}

Nulling specifies constraints on the target Givens rotation $T_{mn}$, which zeroes an element of $X$ by right-multiplication (Fig.~\ref{fig:fs5}).  Assuming unitarity of all matrices, the zeroing of $X_{ij}$ implies that:
\beq
	T_{mn} \begin{bmatrix} -v \\ u \end{bmatrix} = \begin{bmatrix} 0 \\ * \end{bmatrix} \label{eq:si-tuv}
\eeq
i.e.\ the power at the top output is zero when the fields $(-v, u)$ are input to the crossing.  This is equivalent to the splitting-ratio condition $s = \hat{s}$, where $s \equiv (T_{mn})_{11}/(T_{mn})_{12}$ is the splitting ratio of the crossing (Eq.~(\EqII), main text), and $\hat{s} \equiv u/v$ is the target value.  The difficulty in this procedure lies in the difficulty of accurately realizing $\hat{T}_{mn}$ in practice, since the actual transfer matrix is a function of both the control parameters ($\theta, \phi$) and the unknown fabrication imperfections.  Therefore, for the {\it realized} Givens rotation, in general $s \neq \hat{s}$, which will lead to errors in the realized matrix $U$.

There are three distinct variants of the nulling method that accommodate hardware errors to different degrees: (1) an in-silico approach that does not correct errors \cite{S_Reck1994, S_Clements2016}, (2) measurement-assisted nulling, which corrects errors provided that $s$ does not fall within a forbidden region \cite{S_Miller2013b, S_Hamerly2022}, and (3) an improved measurement-assisted method that partially compensates for the ``uncorrectable'' errors arising from unrealizable splitting ratios.

\subsection{In-Silico}
\label{sec:si-e0}

\begin{figure*}[htbp]
\begin{center}
\includegraphics[width=0.85\textwidth]{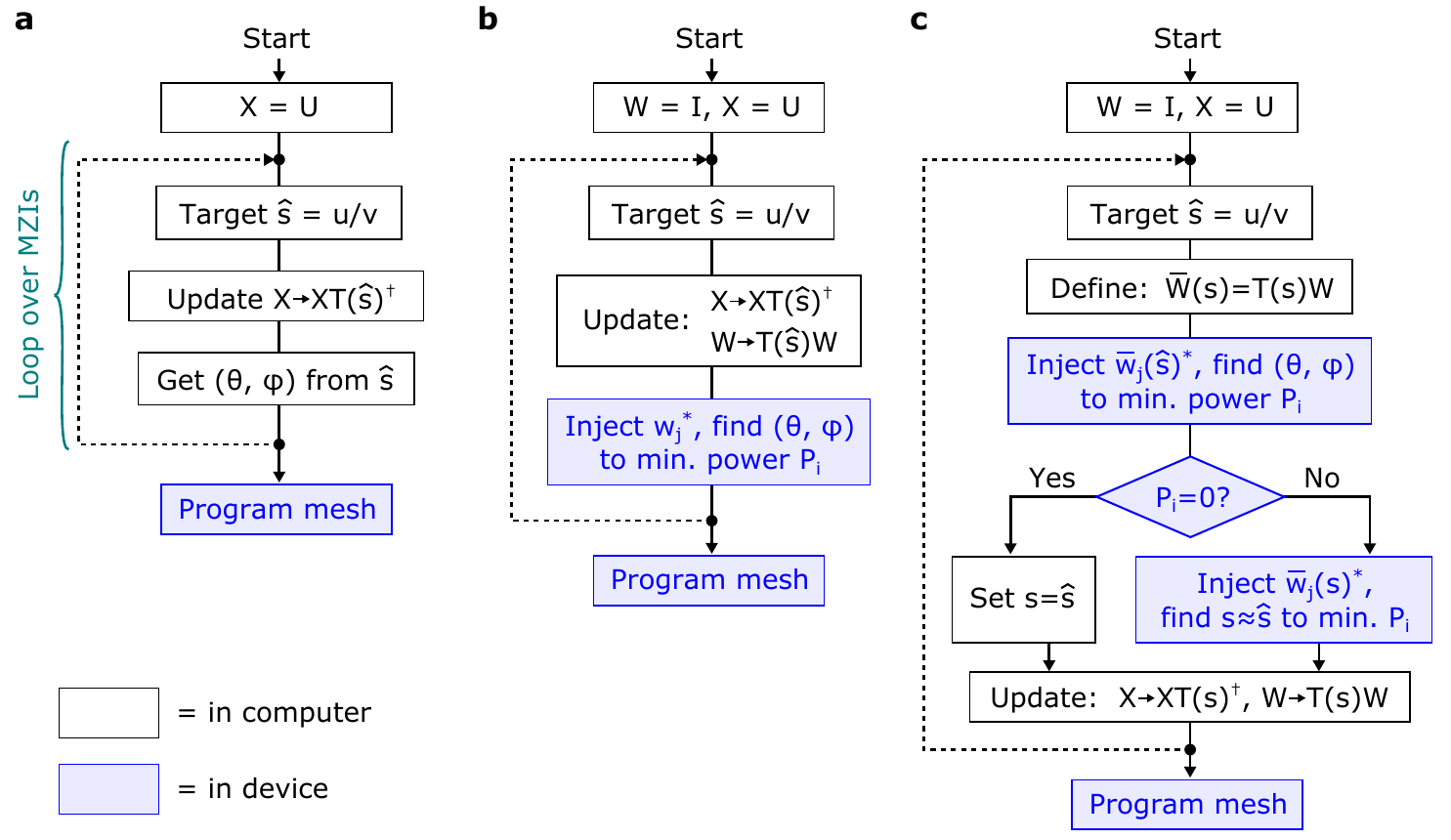}
\caption{Nulling procedure for (a) in-silico (uncorrected) programming \cite{S_Reck1994, S_Clements2016}, (b) measurement-assisted self-configuration \cite{S_Miller2013b, S_Hamerly2022}, and (c) an improvement to the measurement-assisted algorithm.}
\label{fig:fs6}
\end{center}
\end{figure*}

Assuming ideal hardware, there is a simple relation between ($\theta, \phi$) and $T$.  For example, for the standard MZI,
\beq
	T = i e^{i\theta/2} \begin{bmatrix}
		e^{i\phi}\sin(\theta/2) & \cos(\theta/2) \\
		e^{i\phi}\cos(\theta/2) & -\sin(\theta/2)
		\end{bmatrix} \label{eq:si-tideal}
\eeq
in the absence of hardware errors.  Using this formula, we can easily obtain $(\theta, \phi)$ from the target splitting ratio:
\beq
	\theta = 2\tan^{-1} |\hat{s}|,\ \ \ 
	\phi = \text{arg}(\hat{s})
\eeq
Following this procedure, the phase shifts of the mesh are found entirely in a computer.  As a result, hardware errors are not accounted for when programming the mesh, and the realized matrix will be off by an amount called the {\it uncorrected error}:
\beq
	\mathcal{E}_0 \equiv \frac{\langle \lVert U - \hat{U}\rVert \rangle_{\rm rms}}{\sqrt{N}}
	\approx \frac{1}{\sqrt{N}} \sqrt{\sum_{mn} \langle \lVert\Delta T_{mn}\rVert^2 \rangle} \label{eq:si-e0}
\eeq
where $\lVert \cdot \rVert$ is the Frobenius ($L_2$) norm, $U$ and $\hat{U}$ are the realized and target matrices and $\Delta T_{mn} = T_{mn} - \hat{T}_{mn}$ is the difference (due to hardware errors) between the realized $T_{mn}$ and the ideal $\hat{T}_{mn}$ given by Eq.~(\ref{eq:si-tideal}).

In-silico methods were presented in Refs.~\cite{S_Reck1994, S_Clements2016} for the Reck and Clements meshes.  The effect of hardware errors was studied in Refs.~\cite{S_Hamerly2022a, S_Hamerly2022}.  Fig.~\ref{fig:fs6}(a) shows the flowchart for programming a mesh via in-silico nulling.

\subsection{Measurement-Assisted}
\label{sec:si-eco}

In measurement assisted nulling, the first two steps are the same: find the target splitting ratio and update $X$ and $W$ using the corresponding Givens rotation.  The principal difference is that $(\theta, \phi)$ are found using an in-device measurement.  For the Reck mesh, the procedure is traced out in Fig.~\ref{fig:fs4}, where each step attempts to zero a matrix element $X_{ij}$ by injecting $w_j^*$ as input and adjusting the phase shifters to zero the output power at port $i$ (see also Fig.~\ref{fig:fs6}(b)).  This method was first proposed \cite{S_Miller2013b} and demonstrated \cite{S_Annoni2017} on the Reck mesh, but can be generalized to other mesh types provided that tap detectors are present after every MZI \cite{S_Miller2017}.  It was later shown that self-configuration is possible without the tap detectors \cite{S_Hamerly2022}.  Errors occur whenever a crossing cannot be programmed to reach the target splitting ratio, i.e.\ when $\hat{s}$ lies within the forbidden region due to hardware imperfections.  

\subsection{Improved Measurement-Assisted}
\label{sec:si-ecn}

In this paper, we have developed a refinement to the measurement-assisted nulling algorithm that allows for some of the ``uncorrectable'' errors to be partially corrected in subsequent nulling steps.  The impetus for this refinement is the observation that, whenever uncorrectable errors occur, the $s \neq \hat{s}$, and the conventional algorithm as implemented in Fig.~\ref{fig:fs6}(b) incorrectly updates $X$ and $W$.  Error correction can be improved if we can accurately estimate the realized splitting ratio $s$; this allows the algorithm to use this information in order to partially compensate for such errors during the programming of subsequent MZIs.

\begin{table*}[t]
\begin{center}
\begin{tabular}{c|c|c|ccc}
\hline\hline
{\bf Model} & {\bf Arch} & {\bf Coverage} & \multicolumn{3}{c}{{\bf Matrix Error}} \\
& & $\mathcal{C}$ & $\mathcal{E}_0^2$ & $(\mathcal{E}_c^2)_{\rm loc}$ & $(\mathcal{E}_c^2)_{\rm sc}$ \\
\hline
& MZI   & $e^{-N^3 \langle R_+^2 \rangle/24 - N \langle R_-^2\rangle/4}$ 
      	& 
		& $\frac{N^2}{288} \langle R_+^4 \rangle \!+\! \tfrac{1}{48} \langle R_-^4\rangle$
		& \!$\frac{N^2}{432} \langle R_+^4 \rangle \!+\! \frac{\log N-0.422}{24N} \langle R_-^4\rangle$\!
		\\
(any) 
& 3-MZI & $e^{-N(\langle R_+^2\rangle + \langle R_-^2\rangle)}$ 
		& Eq.~(\ref{eq:si-e0}) 
		& $\tfrac{1}{12}(\langle R_+^4\rangle \!+\! \langle R_-^4\rangle)$
		& \!$\frac{\log N-1.366}{3N} (\langle R_+^4\rangle \!+\! \langle R_-^4\rangle)$\!
		\\
& MZI+X & $e^{-N\langle R_+^2\rangle/4 - N^3 \langle R_-^2 \rangle/24}$ 
		& 
		& $\tfrac{1}{48} \langle R_+^4\rangle \!+\! \frac{N^2}{288} \langle R_-^4 \rangle$
		& \!$\frac{\log N-0.422}{24N} \langle R_+^4\rangle \!+\! \tfrac{N^2}{432} \langle R_-^4 \rangle$\!
		\\ 
\hline
& MZI   & $e^{-N^3\sigma^2/3}$ 
		& $2N \sigma^2$
		& $\tfrac23 N^2\sigma^4$
		& $\tfrac49 N^2\sigma^4$
		\\ 
$\sigma$ 
& 3-MZI & $e^{-16N\sigma^2}$
		& $3N\sigma^2$
		& $32\sigma^4$
		& $128 \frac{\log(N)-1.366}{N} \sigma^4$
		\\ 
& MZI+X & $e^{-N^3\sigma^2/3}$
		& $N(2\sigma^2 + \sigma_\gamma^2)$
		& $4\sigma^4$
		& $\tfrac49 N^2\sigma^4$
		\\ 
\hline
& MZI   & $e^{-(2/3)N^3\mu^2}$
		& $4N\mu^2$
		& $\tfrac89 N^2\mu^4$
		& $\tfrac{16}{27} N^2\mu^4$
		\\ 
$\mu$ 
& 3-MZI & $e^{-16N\mu^2}$
		& $3N\mu^2$    
		& $\tfrac{32}{3} \mu^4$
		& $\tfrac{256}{3} \frac{\log N-1.366}{N} \mu^4$
		\\ 
& MZI+X & $e^{-4N\mu^2}$
		& $8(\log(N)-1.422)\mu^2$
		& $\tfrac{16}{3}\mu^4$
		& $\tfrac{32}{3} \frac{\log N-0.422}{N} \mu^4$
		\\ 
\hline
$\mu \gg \sigma$ & (any)
		& $\mathcal{C} = \mathcal{C}_\mu \times \mathcal{C}_\sigma$
		& 
		& $\mathcal{E}^2 = \mathcal{E}^2_\mu + \mathcal{E}^2_\sigma\vphantom{\Bigr|}$ & 
		\\
\hline\hline
\end{tabular}
\caption{Coverage and matrix error for the MZI, 3-MZI, and MZI+X designs.  Matrix error is given for the three nulling methods: in silico (uncorrected, Sec.~\ref{sec:si-e0}), local correction (Sec.~\ref{sec:si-loc}) and self-configuration (Sec.~\ref{sec:si-ecn}).  While the error formulas are general, specific results are given for the uncorrelated model ($\sigma$) and the perfectly correlated model ($\mu$).}
\label{tab:ts4}
\end{center}
\end{table*}

\begin{figure*}[t]
\begin{center}
\includegraphics[width=1.00\textwidth]{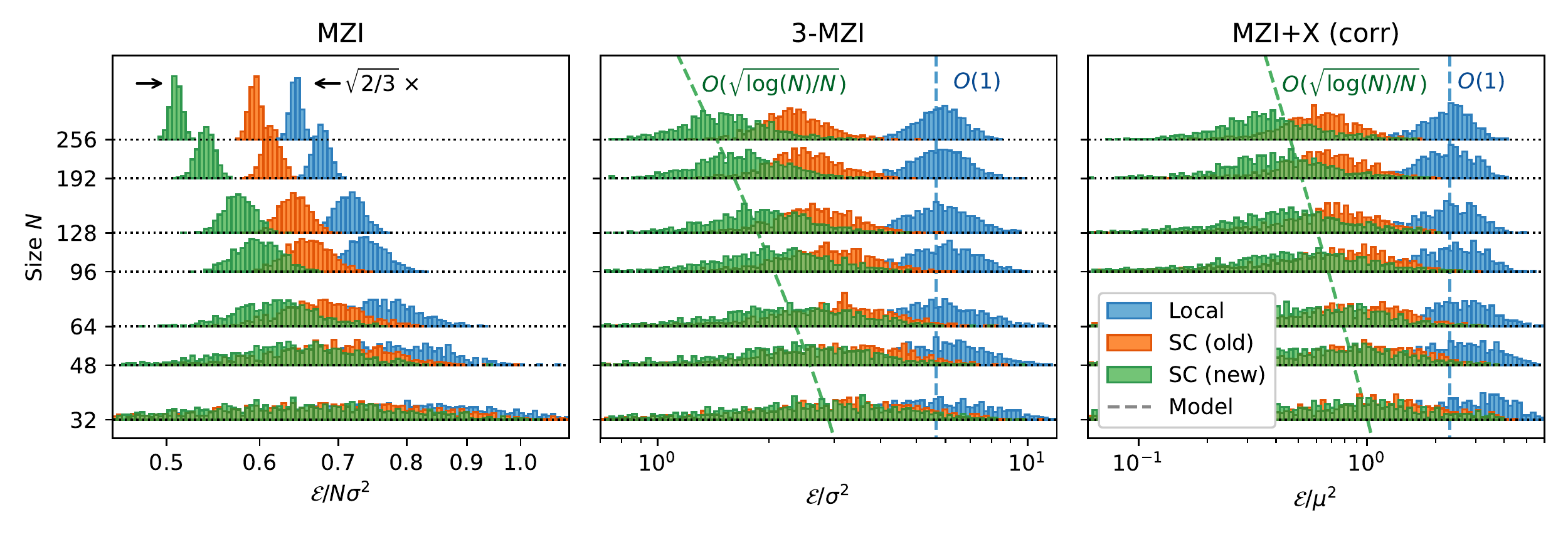}
\caption{Comparison of the accuracy of self-configuration (SC, Sec.~\ref{sec:si-eco}-\ref{sec:si-ecn}) and the local error correction method, Sec.~\ref{sec:si-loc}.  For the MZI and 3-MZI, Gaussian error models are used with $\sigma = 0.05$ and $\sigma = 0.10$, respectively.  For the MZI+X, a correlated error model with $\mu = 0.1$ is used.  Dashed lines correspond to the analytic models in Table~\ref{tab:ts4}.}
\label{fig:fs7}
\end{center}
\end{figure*}

The refined error correction algorithm is shown in Fig.~\ref{fig:fs6}(c).  Here, we defer updates to $X$ and $W$ until the end, and after $(\theta, \phi)$ have been set, we measure $s$ through the following procedure:
\begin{itemize}
	\item If the output power is successfully nulled ($P_i = 0$), then the coupler is configured correctly and $s = \hat{s}$.
	\item If $P_i \neq 0$, nulling is imperfect and $s \neq \hat{s}$.  To find $s$, we now perform an optimization: injecting $\bar{w}_j(s)$ (the $j^{\rm th}$ column of $\bar{W}(s) = T(s)W$, which is a function of $s$), we vary $s$ in the vicinity of $s = \hat{s}$ (with the fixed $(\theta, \phi)$ obtained in the previous step) until the output power is exactly zero.  This procedure obtains the actual splitting ratio implemented in the tunable coupler.
\end{itemize}
Once $s$ is found, we update $X$ and $W$ using $T(s)$.  Since the $W$ and $X$ updates are exact even in the presence of uncorrectable errors, the final matrix error is directly related to the residuals left by imperfect nulling of $X$.  These residuals were calculated in the main text using the formula:
\begin{align}
	& (\mathcal{E}_c^2)_{\rm sc} \nonumber \\
	& = \frac{2}{N} \sum_{mn} \langle r_{mn}^2 \rangle = \frac{1}{2N} \sum_{mn} \underbrace{\bigl\langle|u_{mn}|^2 \!+\! |v_{mn}|^2\bigr\rangle}_{q_{mn}} d(s_{mn}, \hat{s}_{mn})^2 \nonumber \\
	& = \frac{\pi}{12N} \sum_{mn} q_{mn} \bigl[P_{mn}(s_+) \langle R_+^4 \rangle + P_{mn}(s_-) \langle R_-^4 \rangle\bigr] \label{eq:si-ecn}
\end{align}

\subsection{Local Correction Method}
\label{sec:si-loc}

For comparison, we also describe the local method for hardware error correction, first presented in Ref.~\cite{S_Bandyopadhyay2022}.  This method is based on the principle that $2\times 2$ unitary matrices are equivalent up to output phases if they share a common splitting ratio $s \equiv T_{11}/T_{12}$:
\beq
	s = \hat{s} 
	\ \ \Leftrightarrow\ \ 
	T = \begin{bmatrix} e^{i\psi_1} & \\ & e^{i\psi_2} \end{bmatrix} \hat{T} \label{eq:si-loct}
\eeq
This equivalence principle allows perfect MZIs to be substituted for imperfect MZIs columnwise, performing correction locally at each coupler (although the procedure is not strictly local: each step depends on the phases $\psi_i$ accrued from Eq.~(\ref{eq:si-loct}) in the previous step).  Errors occur only when MZI splitting ratios are unrealizable.  These ``uncorrectable errors'' are independent of each other and add up in quadrature.  Refs.~\cite{S_Hamerly2022a, S_Hamerly2022} calculate the resulting matrix error, which follows from the relation
\beq
	\lVert \Delta T \rVert \equiv \text{min}_\psi \Bigl\lVert T - \begin{bmatrix} e^{i\psi_1}\!\! & \\ & \!\!e^{i\psi_2} \end{bmatrix}\hat{T}\Bigr\rVert = \frac{d(s, \hat{s})}{\sqrt{2}}  
\eeq
where $d(s, \hat{s}) = 2|s - \hat{s}| / \sqrt{(|s|^2 + 1)(|\hat{s}|^2 + 1)}$ is the Euclidean metric on the Riemann sphere (under the stereographic projection $s = (x + iy)/(1+z)$, which inverts to $x + iy = 2s/(1+|s|^2)$ and $z = (1-|s|^2)/(1+|s|^2)$).

In the notation of this paper, $\mathcal{E}_c$ takes the form:
\begin{align}
	(\mathcal{E}_{c}^2)_{\rm loc} & \equiv \sum_{mn} \lVert \Delta T_{mn} \rVert^2 = \frac{1}{2N} \sum_{mn} d(s_{mn}, \hat{s}_{mn})^2 \nonumber \\
	& = \frac{\pi}{12N} \sum_{mn} \bigl[P_{mn}(s_+) \langle R_+^4 \rangle + P_{mn}(s_-) \langle R_-^4 \rangle\bigr] \label{eq:si-eco}
\end{align}
Eqs.~(\ref{eq:si-ecn}) and (\ref{eq:si-eco}) are almost identical, differing only by the factor of $q_{mn} = \langle |u_{mn}|^2 + |v_{mn}|^2 \rangle$ in the former.  Since $q_{mn} \leq 1$ due to the unitarity of $X$, Eq.~(\ref{eq:si-ecn}) will always give a lower matrix error.

Table~\ref{tab:ts4} lists the formulas for coverage (Eq.~(\EqXI), main text) and matrix error (Eqs.~(\ref{eq:si-e0}-\ref{eq:si-ecn})) for the three mesh architectures and error models.   We see that, for uncorrelated errors, only the 3-MZI is asymptotically perfect, while both the 3-MZI and MZI+X are asymptotically perfect for correlated errors.  In addition, the uncorrected error can only be reduced in the correlated case, and only for the MZI+X.  Finally, the examples of the 3-MZI and MZI+X highlight the superior performance of the improved self-configuration method.  Under the original method, $\mathcal{E}_c$ is independent of $N$, making the mesh types infinitely scalable (with respect to these errors) but not asymptotically perfect.  But under the improved method, $\mathcal{E}_c \propto \sqrt{\log(N)/N}$, which vanishes in the limit $N \rightarrow \infty$.

Fig.~\ref{fig:fs7} plots the numerically computed accuracy on the three mesh types.  For the 3-MZI and MZI+X, the difference in scaling with $N$ is very clear.  For the regular MZI, all methods give the same scaling, but self-configuration leads to an error lower by a factor of $\sqrt{2/3}$ ($\sqrt{2/3}N\sigma^2$ vs.\ $(2/3)N\sigma^2$).  The overall error amplitude in the figure is distorted by saturation when $\mathcal{E} \sim 1$, but the factor of $\sqrt{2/3}$ is still clearly apparent.

\subsection{Comparison to Global Optimization}
\label{sec:si-glob}

\begin{figure}[tbp]
\begin{center}
\includegraphics[width=1.00\columnwidth]{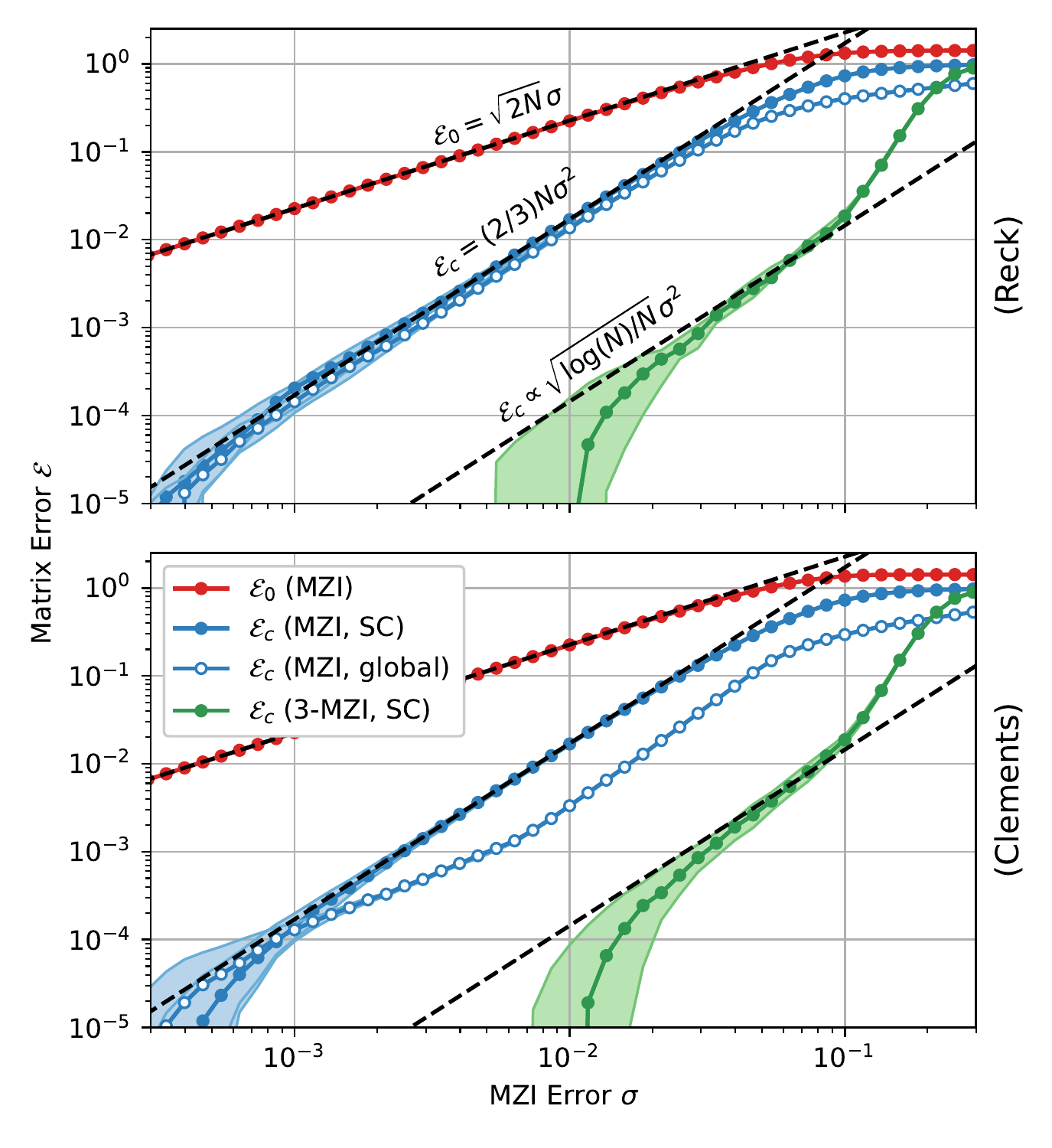}
\caption{Comparison of self-configuration and global optimization for $N = 256$ meshes of the Reck (top) and Clements (bottom) topology, with uncorrelated errors.}
\label{fig:fs8}
\end{center}
\end{figure}

Before the self-configuration and local algorithms were developed, the only way to train imperfect meshes involved global optimization \cite{S_Mower2015, S_Burgwal2017, S_Pai2019}.  Since meshes are linear and reciprocal devices, backpropagation of gradients is equivalent to traversing the mesh in the opposite direction \cite{S_Hughes2018}.  This is implemented in most simulation packages, including \textsc{Neurophox} \cite{S_Neurophox} (based on PyTorch backend) and \textsc{Meshes} \cite{S_Meshes} (based on NumPy with Numba/CUDA extensions), and leads to optimization times orders of magnitude shorter than gradient-free methods.

\begin{figure}[tbp]
\begin{center}
\includegraphics[width=1.00\columnwidth]{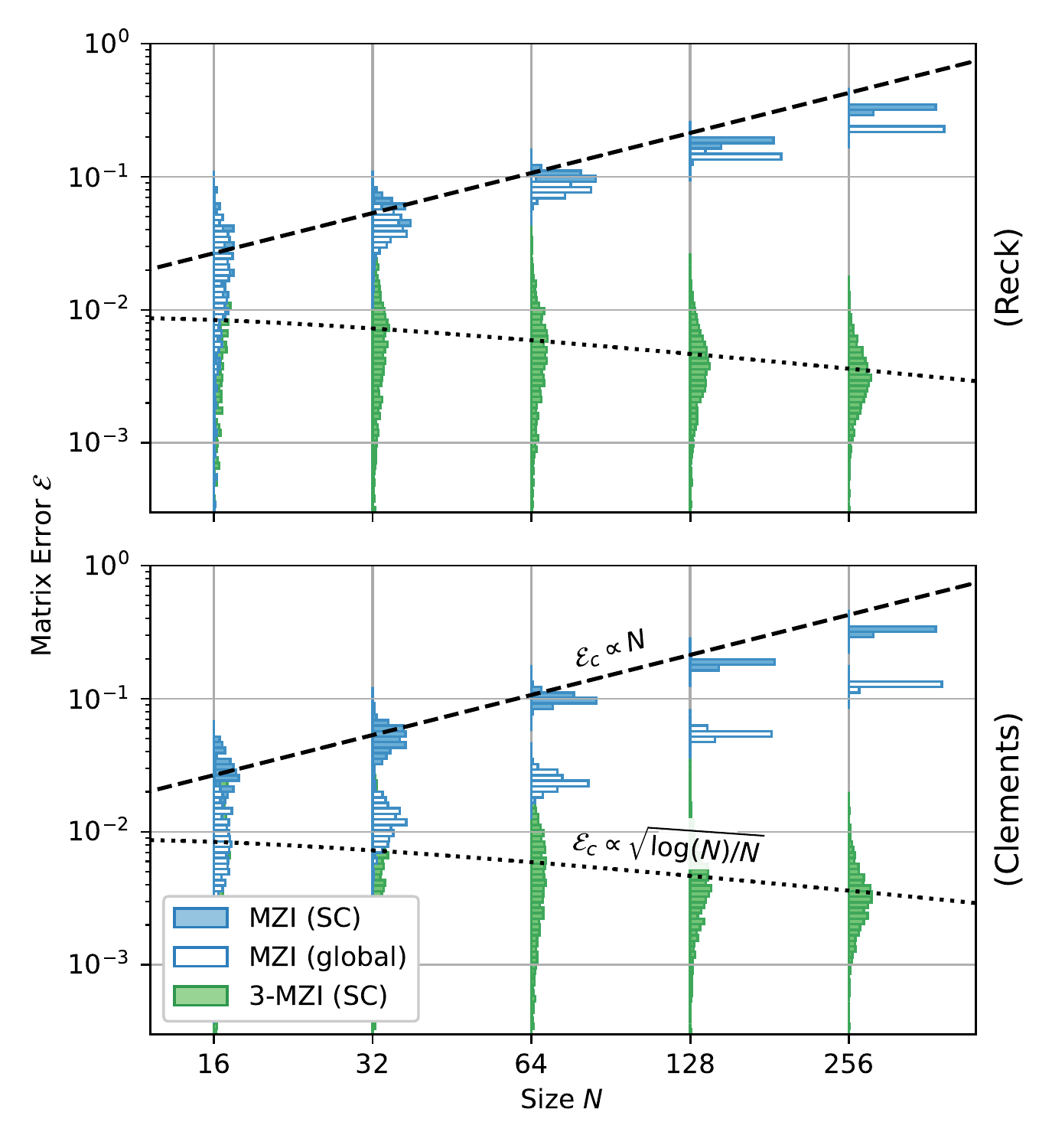}
\caption{Dependence of corrected matrix error $\mathcal{E}_c$ on mesh size.  
Model: uncorrelated splitter errors  with $\sigma = 0.05$.}
\label{fig:fs9}
\end{center}
\end{figure}

Previous studies have shown that gradient-based optimization can give a slight improvement in the matrix fidelity compared to the local or self-configuration approaches \cite{S_Hamerly2022a, S_Hamerly2022}, but take significantly longer to run, in practice requiring thousands of iterations to converge to a solution that is non-negligibly better than the algorithms of Sec.~\ref{sec:si-ecn}.  However, given sufficient computation time, refinement by global optimization may be an appropriate error correction technique.  Using the GPU backend of \textsc{Meshes}, we performed gradient-based optimization on faulty meshes, using the L-BFGS-B algorithm and the self-configured solution as an initial condition.  Figs.~\ref{fig:fs8}-\ref{fig:fs9} compare the accuracy of the self-configured solution to this global refinement.  Interestingly, the improvement is fairly significant (3--4$\times$) for Clements, but negligible for Reck.  We speculate that this discrepancy may be attributed to the triangular structure of Reck, where the MZIs near the apex of the triangle are most likely to lead to uncorrectable errors.  Since the upper-left corner of the matrix depends only on these MZIs, errors in this region cannot be corrected by adjustments to MZIs up- or down-stream.  This is in contrast to the Clements mesh, where all paths pass through an equal number of MZIs, and errors in the center of the mesh (where the probability density clusters close to the cross state) can potentially be corrected by adjustments near the edges.

In both meshes, up to a constant factor, the self-configured and globally-optimized solutions have the same error scaling $\mathcal{E}_c \propto N \sigma^2$ in the MZI mesh.  For large mesh sizes, the 3-MZI mesh still offers a significant improvement over the globally optimized solutions, and its $\mathcal{E}_c \propto \sqrt{\log(N)/N}$ scaling means that this gap grows larger with increasing mesh size.

\section{Imperfectly Correlated Errors}
\label{sec:si-imp}

\begin{figure}[tbp]
\begin{center}
\includegraphics[width=1.00\columnwidth]{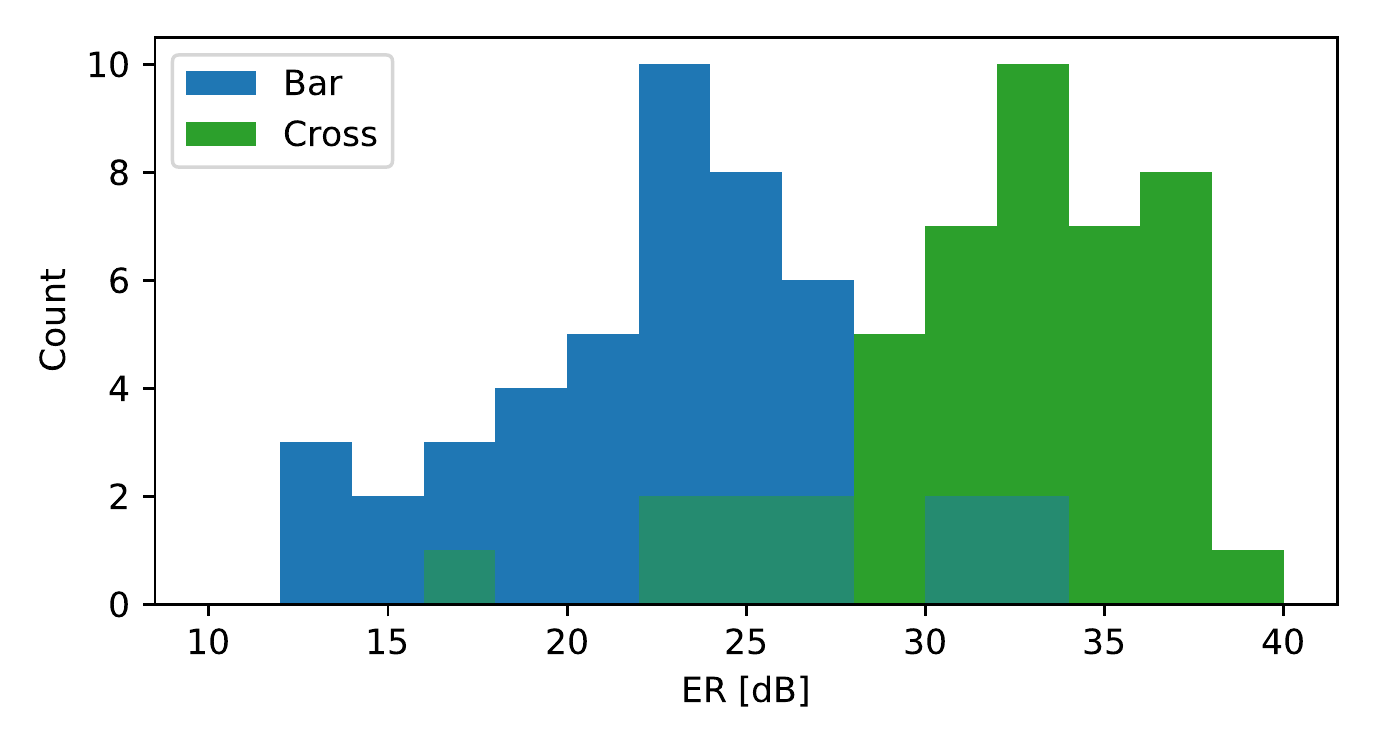}
\caption{Distribution of extinction ratios for MZIs characterized in 3-layer neural network chip of Ref.~\cite{S_Bandyopadhyay2022}.}
\label{fig:fs1}
\end{center}
\end{figure}

The splitter errors $(\alpha, \beta)$ of an MZI are best characterized by measuring the device's extinction ratio.  To do so, one tunes the internal phase shifter $\theta$ and measures the contrast of the interference fringes on the bar- and cross-port outputs.  As an MZI has the following transfer matrix
\begin{align}
	T(\theta, \phi) & = 
		\begin{bmatrix} \cos(\tfrac\pi4+\beta) & i\sin(\tfrac\pi4+\beta) \\ 
			i\sin(\tfrac\pi4+\beta) & \cos(\tfrac\pi4+\beta) \end{bmatrix} 
		\begin{bmatrix} e^{i\theta} & 0 \\ 0 & 1 \end{bmatrix} \nonumber \\
		& \ \ \ \ \times \begin{bmatrix} \cos(\tfrac\pi4+\alpha) & i\sin(\tfrac\pi4+\alpha) \\ 
			i\sin(\tfrac\pi4+\alpha) & \cos(\tfrac\pi4+\alpha) \end{bmatrix} 
		\begin{bmatrix} e^{i\phi} & 0 \\ 0 & 1 \end{bmatrix}
\end{align}
the bar- and cross-port outputs are have extrema $\theta \in \{0, \pi\}$.  The and the extinction ratios are given by:
\begin{align}
	\text{ER}_{\rm bar}\text{[dB]} & = 20 \log_{10} \Bigl| \frac{T_{11}(\theta=\pi)}{T_{11}(\theta\!=\!0)} \Bigr| = 20\log_{10} \Bigl|\frac{\cos(\alpha\!-\!\beta)}{\sin(\alpha\!+\!\beta)}\Bigr| \nonumber \\
	& \approx -20\log_{10} |\alpha+\beta| \\
	\text{ER}_{\rm cross}\text{[dB]} & = 20 \log_{10} \Bigl| \frac{T_{21}(\theta=\pi)}{T_{21}(\theta\!=\!0)} \Bigr| = 20\log_{10} \Bigl|\frac{\cos(\alpha\!+\!\beta)}{\sin(\alpha\!-\!\beta)}\Bigr| \nonumber \\
	& \approx -20\log_{10} |\alpha-\beta|
\end{align}
These relations can be inverted to give us:
\beq
	|\alpha + \beta| = 10^{-\text{ER}_{\rm bar}/20},\ \ \ 
	|\alpha - \beta| = 10^{-\text{ER}_{\rm cross}/20} \label{eq:si-ab}
\eeq
In most photonic platforms, splitter errors are strongly correlated so that $\text{ER}_{\rm cross} \gg \text{ER}_{\rm bar}$.  For example, in Fig.~\ref{fig:fs1}, we plot a histogram of measured MZI extinction ratios characterized for a 3-layer silicon-photonic neural network chip reported in Ref.~\cite{S_Bandyopadhyay2022}.  The median bar- and cross-port extinction ratios are 23~dB and 32~dB, respectively.  This correlation between splitter errors originates from the lengthscales of fabrication process variations that affect the critical dimensions (width, height spacing) of the directional couplers.  These variations typically have correlation lengths on the order of millimeters \cite{S_Yang2015, S_Chrostowski2014, S_Bogaerts2019}, significantly longer than the spacing between couplers in an MZI.  This trend is also observed elsewhere in the literature, as shown in Table~\ref{tab:ts2}.  This suggests an imperfectly-correlated error model of the form

\begin{table}[tbp]
\begin{center}
\begin{tabular}{c|cccc} 
\hline\hline
Ref & Type & Platform & ER$_{\rm cross}$ & ER$_{\rm bar}$ \\ \hline
\cite{S_Kawachi1993} & MZI & SiO$_2$ PLC & 29 & -- \\
\cite{S_Nagase1994} & MZI & SiO$_2$ PLC & 25.9 & -- \\
\cite{S_Goh1999} & MZI & SiO$_2$ PLC & 32.5 & -- \\
\cite{S_Okuno1999} & MZI & SiO$_2$ PLC & 31 & 22 \\
\cite{S_Shoji2010} & MZI & SOI & 35 & 25 \\
\cite{S_Harris2017} & MZI & SOI & 34 & -- \\
\cite{S_Dumais2017b} & MZI & SOI & 34 & 35 \\
\cite{S_Suzuki2018} & MZI & SOI & 41.2 & -- \\
\cite{S_Wilkes2016} & MZI & SOI & -- & 30.9 \\ 
\cite{S_Dong2022} & MZI & SiN:AlN & 30 & -- \\ \hline
\cite{S_Suzuki2015} & Suzuki & SOI & \multicolumn{2}{c}{50.4} \\
\cite{S_Wilkes2016} & Miller & SOI & \multicolumn{2}{c}{60.5} \\
\hline\hline
\end{tabular}
\caption{Reported bar- and cross-port MZI extinction ratios.  Extinction ratios are reported in dB.}
\label{tab:ts2}
\end{center}
\end{table}
\begin{table}[tbp]
\begin{center}
\begin{tabular}{c|ccc}
\hline\hline
& MZI & 3-MZI & MZI+X \\ \hline
$s_+$ & $0$ & $i$ & $\infty$ \\
$s_-$ & $\infty$ & $-i$ & $0$ \\
\hline\hline
\end{tabular}
\caption{Locations of the forbidden regions for each mesh crossing geometry.}
\label{tab:ts1}
\end{center}
\end{table}
\beq
	\alpha \sim N(\mu, \sigma),\ \ \ 
	\beta \sim N(\mu, \sigma)
\eeq
with $\mu \gg \sigma$, is most accurate.  We can use Eq.~(\ref{eq:si-ab}) to relate $\mu$ and $\sigma$ to the median MZI extinction ratios, as follows:
\beq
	\mu = \frac{10^{-\text{ER}_{\rm bar}/20}}{2},\ \ \ 
	\sigma = \frac{10^{-\text{ER}_{\rm cross}/20}}{2.10} 
	\label{eq:si-musig}
\eeq
Recall that the Riemann sphere has two forbidden regions centered at $s_\pm$ (see Table~\ref{tab:ts1}) with radii $R_\pm = 2|\alpha \pm \beta|$.  Under this model, $R_\pm$ has the following moments:
\begin{align}
	\langle R_+^2 \rangle & = 16\mu^2 \bigl(1 + \tfrac12 (\sigma/\mu)^2\bigr)
	& \langle R_-^2 \rangle & = 8\sigma^2 \nonumber \\
	\langle R_+^4 \rangle & = 256\mu^4 \bigl(1 + 3(\sigma/\mu)^2 + \tfrac34(\sigma/\mu)^4\bigr)
	& \langle R_-^4 \rangle & = 192\sigma^4
	\label{eq:si-rpm}
\end{align}
The locations of the forbidden regions are given in Table~\ref{tab:ts1}.  Following the derivation in the Methods (specifically Eqs.~(\EqXII, \EqXVII, \EqXIX, \EqXXIV)), we find:
\beq
	(\mathcal{E}_c)^2 = 
	\begin{dcases} 
		\frac{N^2}{432} \langle R_+^4 \rangle + \frac{\log(N)-0.422}{24N} \langle R_-^4 \rangle & \text{(MZI)} \\
		\frac{\log(N)-1.366}{3N} \bigl(\langle R_+^4 \rangle + \langle R_-^4 \rangle\bigr)  & \text{(3-MZI)} \\
		\frac{\log(N)-0.422}{24N} \langle R_+^4 \rangle + \frac{N^2}{432} \langle R_-^4 \rangle & \text{(MZI+X)}
	\end{dcases}
	\label{eq:si-ec}
\eeq
where we have substituted $(\tfrac54 + \log(2) - \gamma_e) \rightarrow 1.366$ and $(1 - \gamma_e) \rightarrow 0.422$ for clarity.  

\begin{figure}[tb]
\begin{center}
\includegraphics[width=1.00\columnwidth]{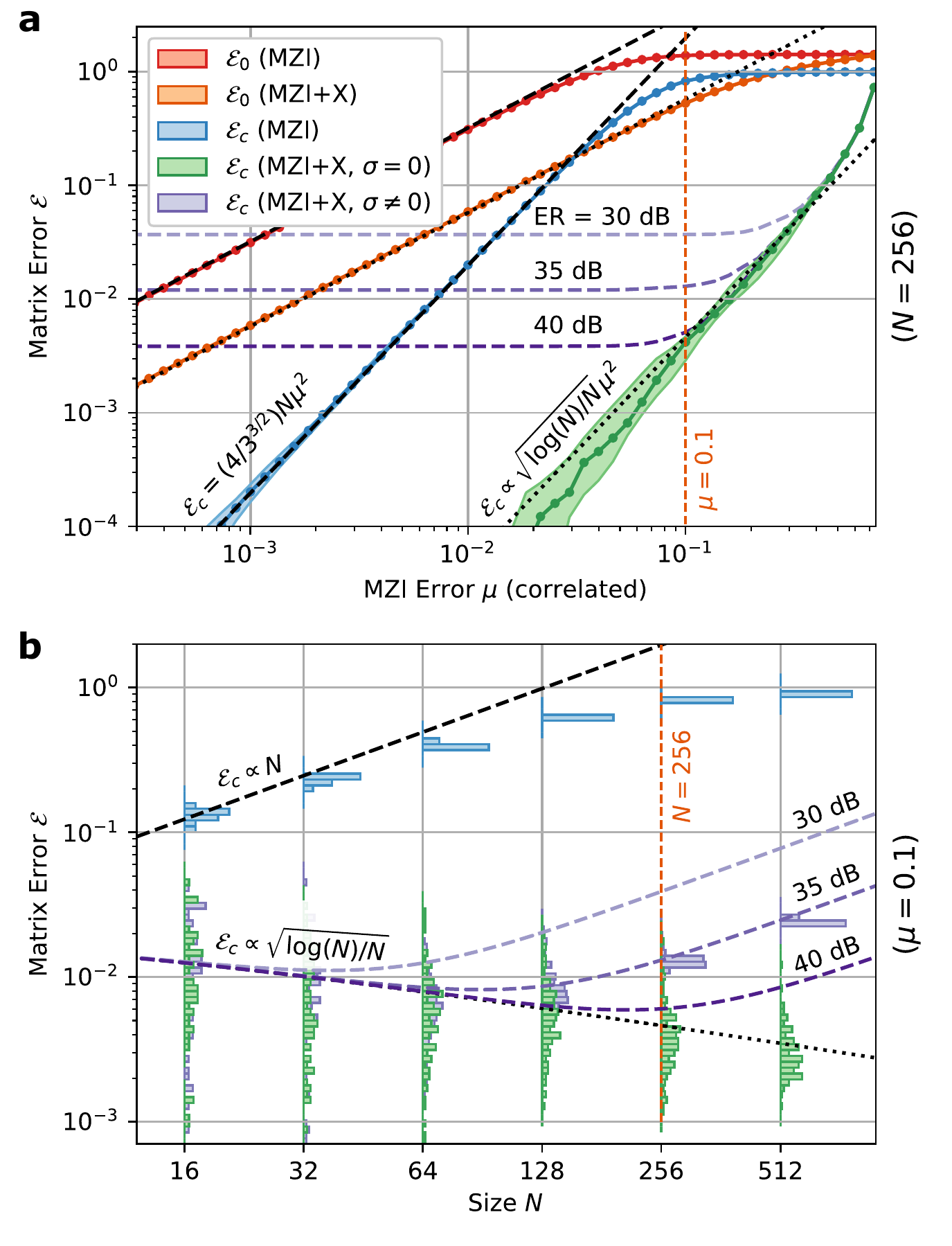}
\caption{Effect of finite cross-port extinction ratio on corrected error for MZI+X; compare Fig.~\FigVI~(main text).  (a) Dependence of matrix error on $\mu$ for a Reck mesh of fixed size $N = 256$.  (b) Dependence on $N$ for fixed $\mu = 0.1$.}
\label{fig:fs2}
\end{center}
\end{figure}

In the main text, we considered the special cases (1) {\it Uncorrelated errors}, $\mu = 0$, where Eqs.~(\ref{eq:si-ec}) reduces to Eqs.~(\EqIII, \EqXII, \EqXX) (main text), and (2) {\it Perfectly correlated errors}, $\sigma = 0$, where Eqs.~(\ref{eq:si-ec}) reduces to Eqs.~(\EqIV, \EqXXV) (main text).  Here we consider the imperfectly correlated case, where $\sigma \ll \mu$.  Substituting Eqs.~(\ref{eq:si-rpm}) into Eqs.~(\ref{eq:si-ec}) and only keeping terms leading order in $(\sigma/\mu)$, we find:
\begin{align}
	& \mathcal{E}_c = \nonumber \\
	& \begin{dcases}
		\frac{4}{3^{3/2}} N \mu^2 & \text{(MZI)} \\
		\frac{16\mu^2}{\sqrt{3}} \Bigl[\frac{\log(N)-1.366}{N}\Bigr]^{1/2} & \text{(3-MZI)} \\
		\Bigl[\frac{32\mu^4}{3} \frac{\log(N)-0.422}{N} + \bigl((2/3)N\sigma^2\bigr)^2\Bigr]^{1/2} & \text{(MZI+X)}
	\end{dcases}
	\label{eq:si-ec2}
\end{align}
For the MZI and 3-MZI, the error is determined entirely by the mean value $\mu$.  On the other hand, for the MZI+X design, the scaling with $N$ in Eq.~(\ref{eq:si-ec2}) means that the accuracy of large meshes is limited by the differential term $\sigma$ even though $\sigma \ll \mu$.  This is shown in Fig.~\ref{fig:fs2}, which shows the effect of nonzero $\sigma$ (characterized in terms of the cross-port ER through Eq.~(\ref{eq:si-musig})) on the MZI+X mesh.  We see that these small differential errors ultimately limit the scaling of this mesh, which is only asymptotically perfect in the ideal case of perfectly correlated errors.  However, for a reasonable value of $\text{ER}_{\rm cross} = 35$~dB (see Table~\ref{tab:ts2}), $\mathcal{E}_c$ is at most a few percent for mesh sizes up to $N = 512$.  This suggests that error correction allows the MZI+X to be asymptotically perfect on all practical mesh sizes, as scaling to meshes of size $N > 512$ is likely prohibitively challenging due to chip area and loss constraints.

In order to exactly cancel the differential term $\alpha - \beta$ as required for very large meshes $N > 1024$, one can place a heater above the directional coupler \cite{S_Orlandi2013}.  While this scheme does come with the cost of an additional active component (putting it in the same complexity category as the ``perfect optics'' approaches \cite{S_Miller2015, S_Suzuki2015}), such an MZI+X with coupler trimming is unique in that it enjoys natively broad bandwidth, enhancing the WDM capacity of the system, which may prove critical to achieving competitive performance in photonic computing applications \cite{S_Feldmann2021}.

\begin{table*}[t!]
\begin{center}
\begin{tabular}{c|c|cc|ccc|ccc|ccc}
\hline\hline
{\bf Ref}	& {\bf Platform} & \multicolumn{2}{c|}{{\bf WG} [$\mu$m]} & \multicolumn{3}{c|}{{\bf Dimensions} [$\mu$m]} & \multicolumn{3}{c|}{{\bf Length Multiplier}} & \multicolumn{3}{c}{{\bf Area Multiplier}}  \\
& & $\ell_{\rm ph}$ & $\ell_{\rm bs}$ & $w_{\rm ph}$ & $w_{\rm bs}$ & $h$ & 3-MZI$^\dagger$ & Suzuki & Miller & 3-MZI$^\dagger$ & Suzuki & Miller \\ \hline
\cite{S_Harris2017} & SOI & 80 & 170 & 80 & 100 & 140 & $1.34\times$ & $1.5\times$ & $2.0\times$ & $1.27\times$ & $1.5\times$ & $2.0\times$ \\
\cite{S_Suzuki2018} & SOI &  80 & 180 & 80 & 80 & 200 & $1.35\times$ & $1.5\times$ & $2.0\times$ & $1.25\times$ & $1.5\times$ & $2.0\times$ \\
\cite{S_Suzuki2015} & SOI & 180 & 200 & 180 & 90 & 180 & $1.26\times$ & $1.5\times$ & $2.0\times$ & $1.17\times$ & $1.5\times$ & $2.0\times$ \\
\cite{S_Wang2020} & SOI & 950 & 130 & 250 & 90 & 130 & $1.06\times$ & $1.5\times$ & $2.0\times$ & $1.13\times$ & $1.5\times$ & $2.0\times$ \\
\cite{S_Wilkes2016} & SOI & 200 & 220 & 200 & 125 & 200 & $1.26\times$ & $1.5\times$ & $2.0\times$ & $1.19\times$ & $1.5\times$ & $2.0\times$ \\
\cite{S_Bandyopadhyay2022} & SOI & 200 & 160 & 200 & 80 & 150 & $1.22\times$ & $1.5\times$ & $2.0\times$ & $1.14\times$ & $1.5\times$ & $2.0\times$ \\
\cite{S_Taballione2022} & SiN & 1300 & 400 & 1300 & 400 & 300 & $1.12\times$ & $1.5\times$ & $2.0\times$ & $1.12\times$ & $1.5\times$ & $2.0\times$ \\
\cite{S_Dong2022} & SiN:AlN & $10^4$ & $10^3$ & 200 & 100 & 1200 & $1.05\times$ & $1.5\times$ & $2.0\times$ & $1.17\times$ & $1.5\times$ & $2.0\times$ \\
\cite{S_Wu2019} & LiNbO$_3$ & $10^4$ & $10^3$ & $10^4$ & $10^3$ & 100 & $1.05\times$ & $1.5\times$ & $2.0\times$ & $1.05\times$ & $1.5\times$ & $2.0\times$ \\
\hline\hline
\end{tabular}
\caption{Waveguide (WG) length $\ell$ and on-chip areal dimensions ($w \times h$) of phase shifters and beamsplitters on several published photonic platforms.  The corresponding unit-cell length and area $A = wh$ (normalized to the standard MZI) are computed from these dimensions.  $^\dagger$MZI+X will have a size similar to 3-MZI.}
\label{tab:ts3}
\end{center}
\end{table*}

\section{Length and Area Estimates}
\label{sec:si-est}

Table~\TabII~of the main text provides a rough comparison of the resource costs of various mesh architectures.  In all cases, the ``perfect optics'' designs \cite{S_Miller2015, S_Suzuki2015} require 1.5--2$\times$ more active components, an important near-term concern as the size of existing chips is often limited by electronic packaging \cite{S_Siew2021} or power dissipation from heaters \cite{S_Kumar2021}.  Waveguide length (which limits loss and SNR \cite{S_AlQadasi2022} and on-chip latency \cite{S_Shen2017, S_Bandyopadhyay2022}) and chip area are also critical parameters, but depend on the implementation.

The approximate MZI dimensions of a range of photonic mesh platforms are reported in Table~\ref{tab:ts3}.  Most SOI devices has similar sizes, although there is a wider range of phase-shifter lengths owing to design tradeoffs (longer thermo-optic phase shifters can be more energy-efficient in certain cases \cite{S_Qiu2020} and the higher heater resistance reduces the required current, but such devices suffer from increased loss and/or higher drive voltages).  Non-SOI platforms such as silicon nitride and lithium niobate can support shorter optical wavelengths and offer mechanisms for faster pure-phase modulation, but suffer from reduced integration density due to the weaker phase-shift mechanisms (e.g.\ Pockels \cite{S_Wu2019} or piezo-optomechanical \cite{S_Dong2022}), which require much longer phase shifters.  In such platforms, the length and area reduction for the 3-MZI is particularly pronounced, as these figures depend primarily on the number of phase shifters and not the number of passive components.

\end{document}